\documentclass[preprints,review,accept,moreauthors,pdftex]{mdpi}
\usepackage{color}
\usepackage[final]{changes}

\firstpage{1} 
\pubvolume{0}
\issuenum{0}
\articlenumber{0}
\pubyear{2023}
\copyrightyear{2023}
\datereceived{} 
\dateaccepted{} 
\datepublished{} 
\hreflink{https://doi.org/} 

\Title{Confronting strange stars with compact-star observations and new physics}

\TitleCitation{Confronting strange stars with compact-star observations and new physics}

\Author{Shu-Hua Yang $^{1}$*\orcidA{}, Chun-Mei Pi$^{2,3}$, Xiao-Ping Zheng $^{1,4}$, and Fridolin Weber $^{5,6}$\orcidB{}}

\AuthorNames{Shu-Hua Yang, Chun-Mei Pi, Xiao-Ping Zheng and Fridolin Weber}

\AuthorCitation{Yang, S.-H.; Pi, C.-M.; Zheng, X.-P; Weber, F.}

\address{%
$^{1}$ \quad Institute of Astrophysics, Central China Normal University, Wuhan 430079, China \\
$^{2}$ \quad School of Physics and Mechanical \& Electrical Engineering, Hubei University of Education, Wuhan 430205, China \\
$^{3}$ \quad Research Center for Astronomy, Hubei University of Education, Wuhan 430205, China \\
$^{4}$ \quad Department of Astronomy, School of physics, Huazhong University of Science and Technology, Wuhan 430074, China \\
$^{5}$ \quad Department of Physics, San Diego State University, San Diego, CA 92182, USA \\
$^{6}$ \quad Center for Astrophysics and Space Sciences, University of California at San Diego, La Jolla, CA 92093, USA
}

\corres{Correspondence: ysh@ccnu.edu.cn}

\abstract{Strange stars ought to exist in the universe according to the strange quark matter hypothesis, which states that matter made of roughly equal numbers of up, down, and strange quarks could be the true ground state of baryonic matter rather than ordinary atomic nuclei. Theoretical models of strange quark matter, such as the standard MIT bag model, the density-dependent quark mass model, or the quasi-particle model, however, appear to be unable to reproduce some of the properties (masses, radii and tidal deformabilities)  of recently observed compact stars. This is different if alternative gravity theory (e.g., non-Newtonian gravity) or dark matter (e.g., mirror dark matter) are considered, which resolve these issues. The possible existence of strange stars could thus provide a clue to new physics, as discussed in this review.}


\begin{document}


\section{Introduction}


The equation of state (EOS) of the dense matter in compact stars (neutron stars, strange stars) is still a mystery today \cite{Mann2020}. The observations of the global properties of compact stars, such as masses and radii, have the potential to unravel this mystery \cite{Ozel2016,Lattimer2016}. In the past few years, tight constraints on the EOS of dense matter were obtained from the tidal deformability derived from the LIGO/Virgo observation of the binary neutron star merger GW170817 and the precise measurements of mass and radius of compact stars by NICER (Neutron star Interior Composition Explorer) \cite{
Raithel2019,Lattimer2019,Orsaria2019,Baiotti2019,Li2019,Li2020,Burgio2020,Chatziioannou2020,Lattimer2021,Li2021}.

Aside from neutron stars (NSs), strange stars (SSs) could also exist in the universe \cite{Farhi1984,Alcock1986,Haensel1986,Alcock1988,Madsen1999,Weber2005}, which are made of strange quark matter (SQM) consisting of up, down and strange quarks. This is a consequence of the hypothesis proposed by Itoh \cite{Itoh1970}, Bodmer \cite{Bodmer1971}, Witten \cite{Witten1984}, and Terazawa \cite{Terazawa1989a,Terazawa1989b} which states that SQM could be the true ground state of baryonic matter rather than conventional atomic nuclei. The galaxy is likely to be contaminated by strange quark nuggets if SQM is the true ground state, and NSs could be converted to SSs by these strange quark nuggets \cite{Madsen1999,Glendenning1990,Caldwell1991}. Therefor, many compact stars (neutron stars) could actually be SSs \cite{Weber2005}.

Many \replaced{researchers}{works} have tried to identify SSs through observations of compact stars (see \cite{Weber2005,Bhattacharyya2016}, and references therein). For example, Di Clemente et al. \cite{Di Clemente2022} \added{and Horvath et al. \cite{Horvath2023}} suggest that the central compact object within the supernova remnant HESS J1731-347 could be \replaced{an}{a} SS rather than \replaced{an}{a} NS because of its small mass (a mass of the order or smaller than one solar mass) \cite{Doroshenko2022}. The reason is that the analysis of various types of \replaced{supernova}{SN} explosions indicates that it is not possible to produce \replaced{an}{a} NS with a mass smaller than about $1.17\, M_{\odot}$ \cite{Suwa2018}, but it is possible to produce \replaced{an}{a} SS with a small mass. \added{In fact, the mass and radius observations of compact stars have long been used to identify SSs, and some SS candidates have been reported, such as compact stars in the X-ray sources SAX J1808.4-3658 \cite{Li1999a}, 4U 1728-34 \cite{Li1999b}, RX J1856.5-3754 \cite{Drake2002,Burwitz2003}, 4U 1746-37 \cite{Li2015} and the radio pulsar PSR B0943+10 \cite{Yue2006}.}

Besides the possible absolute stability of SQM and the existence of SSs (which is the case for this review), there are other possibilities widely \replaced{investigated}{invesgatied}: 
\begin{itemize}
\item	If SQM is not the true ground state of baryonic matter, \added{but} is only a metastable state, hybrid stars (NSs consisting of \replaced{an}{a} SQM core or a hadron-quark mixed phase) might exist \cite[e.g.,][]{Glendenning2000,Blaschke2018,Annala2020}.
\item   There is the possiblity that metastable hadronic stars could coexist with SSs, which is called the two-families scenario \cite[e.g.,][]{Berezhiani2003,Drago2004,Bombaci2004,Bombaci2021,Bombaci2022}.
\item	Compact stars (neutron stars) might be strangeon stars made of strangeons (coined by combining ‘strange nucleon’) \cite[e.g.,][]{Xu2003,Miao2022a,Lai2023}.
\item	Instead of SSs, up-down quark stars might exist because quark matter made of up and down quarks could be more stable than ordianary nuclear matter and SQM in some models \cite[e.g.,][]{Holdom2018,Zhang2019,Zhang2020,Cao2022}.\end{itemize}

The parameters of the SQM model could be constrained by the observations of compact stars \cite{Weissenborn2011,Wei2012,Pi2015,Zhou2018,Yang2020,Cai2021,Backes2021,Yang2021a,Yang2021b,Pi2022a,Pi2022b}.  For the MIT bag model, Weissenborn et al. \cite{Weissenborn2011} found that the parameters of the model could be constrained if one demands that the maximum mass of SSs must be greater than the mass of PSR J1614-2230 ($1.97 \pm 0.04\, M_{\odot}$ \cite{Demorest2010}). The parameters of the model was further constrained by Zhou et al. \cite{Zhou2018} using both the mass of PSR J0348+0432 ($2.01 \pm 0.04\, M_{\odot}$ \cite{Antoniadis2013}) and the tidal deformability of GW170817 \cite{Abbott2017}. It was found that SSs could exist in certain ranges of the values of the parameters of the SQM model. However, $\Lambda(1.4)\leq800$ \cite{Abbott2017} [$\Lambda(1.4)$ is the dimensionless tidal deformability for a $1.4\, M_{\odot}$ star] was used in that paper, which was updated to $\Lambda(1.4)=190_{-120}^{+390}$ [we will use $\Lambda(1.4)\leq 580$ in this review] \cite{Abbott2018}. Moreover, the largest observed mass of pulsars was updated to $2.14_{-0.09}^{+0.10}\, M_{\odot}$ (PSR J0740+6620) in 2020 \cite{Cromartie2020}. With these new data, Yang et al. \cite{Yang2020} found that the existence of SSs seems to be ruled out if the standard MIT bag model of SQM is used. \replaced{Note}{Mention} that the above conclusion remains correct \cite{Yang2021a} although the mass of PSR J0740+6620 was updated to $2.08 \pm 0.07\, M_{\odot}$ \cite{Fonseca2021}. In addition to the standard MIT bag model, SSs are ruled out by the observations of compact stars for the density-dependent quark mass model \cite{Yang2021b} and the quasi-particle model \cite{Cai2021}.

However, Yang et al. \cite{Yang2020} found that SSs cannot be ruled out if non-Newtonian gravity effects are considered \footnote{For the standard MIT model, this is ture considering the constraints from the observations of PSR J0740+6620 and GW170817, which are mentioned in the last paragraph. However, if the observation data of PSR J0030+0451 are also considered, SSs will be ruled out even if we consider the effects of non-Newtonian gravity \cite{Pi2022a}. The details can be seen in Sect.~\ref{allparanon} of this review.}.
Moreover, aside from the non-Newtonian gravity effects, Yang et al. \cite{Yang2021a} found that the observations of compact stars could be satisfied if a mirror-dark-matter (MDM) core exists in some SSs.

Non-Newtonian gravity is one kind of the alternative theories of gravity, which are beyond General Relativity (GR). \deleted{While, }MDM is one possible candidate of the dark matter (DM), which is beyond the Standard Model (SM) of particles. Both the alternative theory of gravity and DM constitute the new physics that is being vividly discussed today.

Compact stars (NSs or SSs) are dense objects with strong gravity. Although GR agrees well with the experiments in the solar system, it is not fully tested in the strong-field domain \cite{Doneva2018,Li2019}. Thus, compact stars are ideal places to test gravity theories. The properties of compact stars in the framework of the alternative theory of gravity are studied extensively \cite[e.g.,][]{Krivoruchenko2009,Wen2009,Cooney2010,Arapoglu2011,Pani2011a,Pani2011b,Rahaman2012,Harko2013,Staykov2014,Sham2014,Moraes2016,Yagi2017,Yazadjiev2018,Lopes2018,Debabrata2019,Salako2020,Majid2020,Astashenok2020,Saffer2021,Banerjee2021,Danchev2021,Astashenok2021,
Prasetyo2021,Panotopoulos2021,Xu2022,Hanafy2022,Tangphati2022,Pretel2022,Shao2022,Carvalho2022,Lin2022,Yang2022c,Jimenez2022a,Maurya2023}, and for recent reviews, see Refs. \cite{Li2019,Shao2019,Olmo2020}. 

The nature of DM is still unknown today. If DM is self-interacting (but has a negligible annihilation rate) \cite{Spergel2000,Tulin2018, Bertone2018}, compact stars might contain a DM core (or a DM halo), which will impact the global properties of the stars. On the other hand, the observations of compact stars might help us to reveal the nature of DM. The properties of compact stars with a DM core (or a DM halo) are widely studied \cite[e.g.,][]{
Sandin2009,Ciarcelluti2011,Leung2011,Li2012,Li2012a,Xiang2014,Mukhopadhyay2017,Ellis2018,
Deliyergiyev2019,Bezares2019,Ivanytskyi2020,Kain2021,Berezhiani2021,Ciancarella2021,Hippert2022,Gleason2022,Leung2022,Das2022a,Giovanni2021,Karkevandi2022,Emma2022,Miao2022b,Collier2022,Dengler2022,  
Rutherford2022,Shakeri2022,Giangrandi2022,Fynn2023,
Mukhopadhyay2016,Panotopoulos2017,Panotopoulos2018,Jimenez2022b,Ferreira2022,Lopes2023}, and recent reviews can be seen in the introductory part of Ref. \cite{Karkevandi2022}. 


This review is organized as follows: 
In Sect.~\ref{eossqm}, we briefly review the EOS of SQM employed in this paper (i.e., the standard MIT bag model). In Sect.~\ref{sswithout}, we study the structure and the dimensionless tidal deformability of SSs without the consideration of the new physics, and show that in this case, SSs are ruled out by the observations of compact stars. Then, SSs with the non-Newtonian gravity effects are studied in Sect.~\ref{ssnon}, and SSs with a MDM core are studied in Sect.~\ref{ssmdm}. Finally, the conclusions are given in Sect.~\ref{conclusions}.

\section{EOS of SQM} \label{eossqm}


In this review, we use the standard MIT bag model for SQM \cite{Farhi1984,Haensel1986,Alcock1986,Weber2005}. The mass of $u$ and $d$ quarks is taken to be zero\deleted{ in that model}, while the mass of $s$ quarks is nonzero (both $m_s=93$ MeV and $m_s=95$ MeV are considered in this review \cite{Zyla2020} \footnote{For the choice of the value of $m_s$, we just follow the  early papers related to this review. The results are similiar whether we choose $m_s=93$ MeV or $m_s=95$ MeV. In fact, the results will not be changed qualitatively even we choose $m_s=150$ MeV, as shown in Refs. \cite{Yang2020,Yang2021a}.}). First-order perturbative corrections in $\alpha_{S}$ (the strong interaction coupling constant) are considered\deleted{ in that model}.

The thermodynamic potentials for each species of the quarks and the electrons are given by \cite{Alcock1986,Yang2020,Yang2021a}

\begin{equation}
\Omega_{u}=-\frac{\mu_{u}^{4}}{4\pi^{2}}\bigg(1-\frac{2\alpha_{S}}{\pi}\bigg),
\end{equation}

\begin{equation}
\Omega_{d}=-\frac{\mu_{d}^{4}}{4\pi^{2}}\bigg(1-\frac{2\alpha_{S}}{\pi}\bigg),
\end{equation}

\begin{eqnarray}
\nonumber
\Omega_{s}&=&-\frac{1}{4\pi^{2}}\bigg\{\mu_{s}\sqrt{\mu_{s}^{2}-m_{s}^{2}}
(\mu_{s}^{2}-\frac{5}{2}m_{s}^{2})+\frac{3}{2}m_{s}^{4}f\\ \nonumber
&&-\frac{2\alpha_{S}}{\pi}\bigg[3\bigg(\mu_{s}\sqrt{\mu_{s}^{2}-m_{s}^{2}}
-m_{s}^{2}f\bigg)^{2}-2(\mu_{s}^{2}-m_{s}^{2})^{2}-3m_{s}^{4}\textrm{ln}^{2}\frac{m_{s}}{\mu_{s}}\\
&&+6\textrm{ln}\frac{\sigma}{\mu_{s}}\bigg(\mu_{s}m_{s}^{2}\sqrt{\mu_{s}^{2}
	-m_{s}^{2}}-m_{s}^{4}f\bigg)\bigg]\bigg\},
\label{OmegaS}
\end{eqnarray}

\begin{equation}
\Omega_{e}=-\frac{\mu_{e}^{4}}{12\pi^{2}},
\end{equation}
where
$f\equiv\textrm{ln}[(\mu_{s}+\sqrt{\mu_{s}^{2}-m_{s}^{2}})/m_{s}]$, $\sigma$ is a renormalization constant which is of
the order of the chemical potential of $s$ quarks. In this review, we take $\sigma=300$ MeV.

The energy density and the pressure are given by
\begin{equation}
\epsilon_{Q}=\sum_{i=u,d,s,e}(\Omega_{i}+\mu_{i}n_{i})+B,
\label{eq:epsQ}
\end{equation}
\begin{equation}
p_{Q}=-\sum_{i=u,d,s,e}\Omega_{i}-B,
\label{eq:pQ}
\end{equation}
where $B$ is the bag constant, $\mu_{i}$ ($i=u,d,s,e$) are the chemical potentials, and $n_{i}$ are the number densities
\begin{equation}
n_{i}=-\frac{\partial\Omega_{i}}{\partial\mu_{i}}.
\end{equation}
To claculate $\epsilon_{Q}$ and $p_{Q}$, the condition of chemical equilibrium 
\begin{equation}
\mu_{d} =  \mu_{s} =  \mu_{u}+\mu_{e},
\end{equation}
and  electric charge neutrality condition
\begin{equation}
\frac{2}{3}n_{u}-\frac{1}{3}n_{d}-\frac{1}{3}n_{s}-n_{e}=0,
\end{equation}
should be employed.

\section{SSs without the consideration of the new physics} \label{sswithout}

\subsection{The global properties of SSs}\label{tid}

In the following of this review, geometrized units $G=c=1$ are used.

The structure of SSs can be calculated by solving the Tolman-Oppenheimer-Volkoff (TOV) equations (which are for static stars, and are derived in the framework of GR) \cite{Oppenheimer1939,Tolman1939}:
\begin{equation}
\frac{dp(r)}{dr}=-\frac{[m(r)+4\pi r^{3}p(r)][\epsilon(r)+p(r)]}{r[r-2m(r)]},
\end{equation}
\begin{equation}
\frac{dm(r)}{dr}=4\pi \epsilon(r) r^{2}.
\end{equation}
For a given EOS, and for a given pressure at the center of the star, these equations can be solved using the following boundary conditions: $p(R)=0$, $m(0)=0$ .

The definition of the dimensionless tidal deformability is $\Lambda\equiv\lambda/M^{5}$, where $\lambda$ is the tidal deformability parameter \footnote{$\lambda$ symbolizes the tidal deformability parameter here. It also symbolizes the length scale of non-Newtonian gravity conventionally in Eqs.\ (\ref{vr}) and (\ref{mulam}).}. Considering that $\lambda=\frac{2}{3}k_{2}R^{5}$ ($k_{2}$ is the dimensionless tidal Love number and $R$ is the radius) \citep{Flanagan2008,Hinderer2008,Damour2009,Hinderer2010}, we can get
\begin{equation}
\Lambda=\frac{2}{3}k_{2}\beta^{-5},
\label{dtd}
\end{equation}
where $\beta$ ($\equiv M/R$) is the compactness of the star.

One can calculate the tidal Love number $k_{2}$ using the following expression \cite{Lattimer2016}
\begin{equation}
k_{2}=\frac{8}{5}\frac{\beta^{5}z}{F},
\end{equation}
where
\begin{eqnarray}
z=(1-2\beta)^{2}[2-y_{R}+2\beta(y_{R}-1)],
\label{z_w}
\end{eqnarray}
\begin{eqnarray}
\nonumber
F&=&6\beta(2-y_{R})+6\beta^{2}(5y_{R}-8)+4\beta^{3}(13-11y_{R}) \\
&&  +4\beta^{4}(3y_{R}-2)+8\beta^{5}(1+y_{R})+3z\textrm{ln}(1-2\beta).
\label{f_w}
\end{eqnarray}

\replaced{Note}{Mention} that in Eqs.\ (\ref{z_w}) and (\ref{f_w}), $y_{R}$ does not equal to the value of $y(r)$ at the surface of the star [$y(R)$]. In fact, $y_{R} = y(R)-4\pi
R^{3}\epsilon_{s}/M$, because the energy density $\epsilon_{s}$ is nonzero inside the surface of SSs \citep{Postnikov2010}. The quantity $y(r)$ satisfies 
\begin{equation}
\frac{dy(r)}{dr}=-\frac{y(r)^{2}}{r}-\frac{y(r)-6}{r-2m(r)}-rQ(r),
\label{yr_w}
\end{equation}
where
\begin{eqnarray}
\nonumber Q(r)&=&4\pi\frac{[5-y(r)]\epsilon(r)+[9+y(r)]p(r)
	+\frac{\epsilon(r)+p(r)}{\partial p(r)/\partial \epsilon(r)}}{1-2m(r)/r}\\ &&-4\bigg[\frac{m(r)+4\pi
	r^{3}p(r)}{r[r-2m(r)]}\bigg]^{2}.
\label{Qr_w}
\end{eqnarray} 
One can \replaced{solve}{calculate} Eq.\ (\ref{yr_w}) together with the TOV equations for a given EOS, and the boundary condition is $y(0)=2$.

\begin{figure}[H]
\includegraphics[width=10.5 cm]{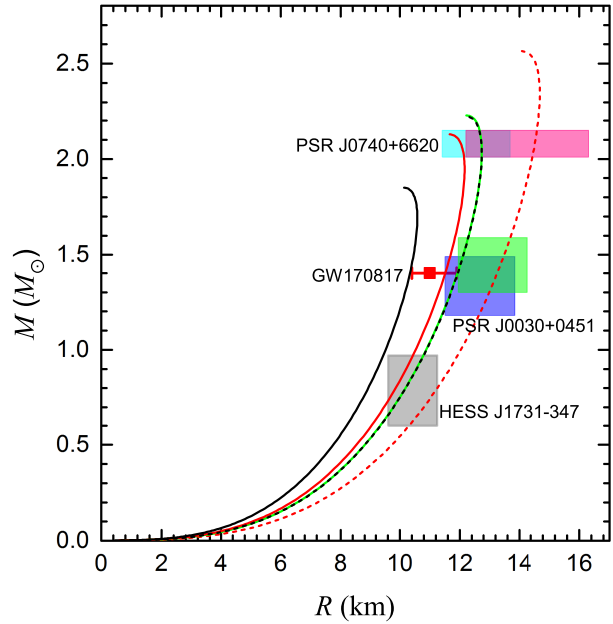}
\caption{The mass-radius relation of SSs for $m_{s}=93$ MeV. The black lines are for $\alpha_{S}=0.4$ (the solid one is for $B^{1/4}=148.4$ MeV and the dashed one is for $B^{1/4}=134.8$ MeV), the red lines are for $\alpha_{S}=0.7$ (the solid one is for $B^{1/4}=138.1$ MeV and the dashed one is for $B^{1/4}=125.1$ MeV), and the green line is for the data set ($\alpha_{S}=0.7$, $B^{1/4}=134.8$ MeV). The red data \added{point} corresponds to the radius of a $1.4\, M_{\odot}$ compact star obtained from the observations of GW170817 \cite{Capano2020}. The blue and green regions correspond to the mass and radius of PSR J0030+0451 obtained from NICER data, which are given by Riley et al. \cite{Riley2019} and Miller et al. \cite{Miller2019}, respectively. The cyan and pink regions are for PSR J0740+6620, where the mass is from Ref. \citep{Fonseca2021} and the radius is obtained from NICER and XMM-Newton data by Riley et al. \cite{Riley2021} and Miller et al. \cite{Miller2021}, respectively. The grey region corresponds to the central compact object within the supernova remnant HESS J1731-347 \cite{Doroshenko2022}.\label{rm}}
\end{figure}   

The mass-radius relation of SSs is shown in Fig.\ \ref{rm}, which is calculated by solving the TOV equations using the standard MIT bag model. All the parameter sets of [$B^{1/4}$(MeV), $\alpha_{S}$] can satisfy both the "2-flavor line" and  "3-flavor line" constraints (see Fig.\ \ref{constraints}). From Fig.\ \ref{rm}, we find that for the fixed value of $\alpha_{S}$, both the maximum mass of SSs and the radius of a 1.4$M_{\odot}$ SS increases significantly with the decreasing of $B^{1/4}$. While, for the fixed value of $B^{1/4}$, these properties change slightly for different values of $\alpha_{S}$ (see the green line and the black dashed line in Fig.\ \ref{rm}).

As shown in Fig.\ \ref{rm}, the red solid line, the green line and the black dashed line can comply with all observed data. But this result turns out to be not true as we will show later in Sect.~\ref{constnon}. The reason is that the data of GW170817 shown in this figure is from Ref. \cite{Capano2020}, which is not based on the study of SSs or the standard MIT bag model. 

\subsection{Parameter space of SQM}\label{constnon}

\begin{figure}[H]
\begin{adjustwidth}{-\extralength}{0cm}
\centering
\includegraphics[width=13.5cm]{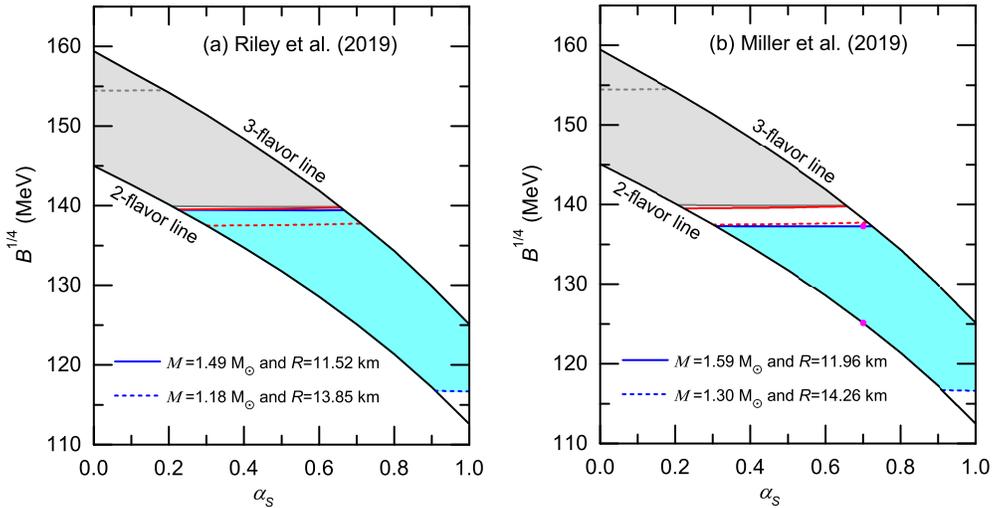}
\end{adjustwidth}
\caption{Constraints on the parameters of the standard MIT model for $m_{s}=93$ MeV. The grey solid and dashed lines are for $\Lambda(1.4)=580$ and $\Lambda(1.4)=190$, respectively. The red solid and dashed lines are for $M_{\rm max} = 2.08\, M_{\odot}$ and $M_{\rm max} = 2.14\, M_{\odot}$, respectively. The magenta dots in (b) correspond to (125.1,0.7) and (137.3, 0.7), which will be used later in Fig.\ \ref{MDM_fig1} and Fig.\ \ref{MDM_fig2}. Image from Ref. \cite{Yang2021a}.\label{constraints}}
\end{figure}

The allowed parameter space of the standard MIT model can be calculated with the following constraints \cite{Schaab1997,Weissenborn2011,Wei2012,Pi2015,Zhou2018,Yang2020,Cai2021,Backes2021,Yang2021a,Yang2021b,Pi2022a}:

First, pure SSs could exist only if SQM is the true ground state of baryonic matter, which means that the energy per baryon of SQM must be smaller than the one of $^{56}$Fe ($E/A\sim 930$ MeV). It is common to compare the energy per baryon of SQM to $^{56}$Fe although the latter is only the third lowest after $^{62}$Ni and $^{58}$Fe. The parameter regions satisfy this constraint are the areas below the 3-flavor line in Fig.\ \ref{constraints}.

The second constraint follows that the energy per baryon of the non-strange quark matter (quark matter made of $u$ and $d$ quarks) in bulk must be larger than $934$ MeV (the additional 4 MeV comes from the correction of the surface effects \cite{Farhi1984,Madsen1999,Weissenborn2011,Zhou2018}). This constraint ensures that atomic nuclei do not dissolve into their constituent quarks. The parameter regions satisfy this constraint are the areas above the 2-flavor line in Fig.\ \ref{constraints}. \added{Here, a small value of the surface tension [$\sigma$=(70 MeV)$^3$ ] was employed \cite{Farhi1984}. However, the exact surface tension value is still uncertain and it could be much larger \cite{Alford2001,Oertel2008,Lugones2013,Xia2013,Lugones2019,Fraga2019,Wang2021}. Therefore, the critical value of $934$ MeV could be smaller, and the 2-flavor line in Fig.\ \ref{constraints} will shift downward, it will leads to a larger allowed parameter region.}  

The two constraints mentioned above must be fulfilled as we dicuss other constraints in the following. In other words, we are only interested in the areas between the 3-flavor line and the 2-flavor line.

The third constraint is $\Lambda(1.4)\leq 580$ [$\Lambda(1.4)$ is the dimensionless tidal deformability of a $1.4\, M_{\odot}$ star], which follows from the observation of GW170817 [$\Lambda(1.4)=190 _{-120}^{+390}$] \cite{Abbott2018}. This constraint leads to the grey-shadowed areas in Fig.\ \ref{constraints}.

The fourth constraint is $M_{\rm max} \geq
2.08\, M_{\odot}$, where $M_{\rm max}$ is the maximum mass of SSs derived from TOV equations, and $2.08\, M_{\odot}$ is the mass of PSR J0740+6620 \cite{Fonseca2021} \footnote{A new massive compact star with a mass of $2.35 \pm 0.17\, M_{\odot}$ is reported recently (PSR J0952-0607) \citep{Romani2022}. It is one of the fastest-spinning pulsars with a spin period of 1.41 ms, and  the rotation effects on the mass and radius cannot be ignored for this star \cite{Konstantinou2022}.}.  The parameter regions satisfy this constraint are the areas below the red solid lines in Fig.\ \ref{constraints}. Since the mass of PSR J0740+6620 was first reported to be $2.14_{-0.09}^{+0.10}\, M_{\odot}$ \cite{Cromartie2020}, we also use $M_{\rm max} \geq 2.14\, M_{\odot}$ in Sect.~\ref{allparanon} following our previous papers \cite{Yang2020,Yang2021b,Pi2022a}. The red dashed lines for $M_{\rm max} = 2.14\, M_{\odot}$ are also shown in Fig.\ \ref{constraints}. As can be seen from Fig.\ \ref{constraints}, both for $M_{\rm max} \geq 2.08\, M_{\odot}$ and $M_{\rm max} \geq 2.14\, M_{\odot}$, one can reach the same conclusion that SSs are ruled out by the observations of PSR J0740+6620 and GW170817.

The last constraint is from the observed mass and radius of PSR J0030+0451. Two independent data are derived from the NICER observations, namely, $M = 1.34_{-0.16}^{+0.15}\, M_{\odot}$ and $R_{\rm eq} = 12.71_{-1.19}^{+1.14}$ km by Riley et al.  \cite{Riley2019}, and $M =1.44_{-0.14}^{+0.15}\, M_{\odot}$ and $R_{\rm eq}=13.02_{-1.06}^{+1.24}$ km by Miller et al. \cite{Miller2019}. These data are translated into the $B^{1/4}$--$\alpha_{S}$ space and shown in Fig.\ \ref{constraints} (see the blue lines). The cyan-shadowed areas in Fig.\ \ref{constraints} correspond to the parameter space allowed by this constraint.

In fact, as can be seen from Fig.\ \ref{constraints}, the cyan-shadowed areas can satisfy not only the constraint from the observation of PSR J0030+0451, but also the constraint from the observation of PSR J0740+6620. We find that the cyan-shadowed area and the grey-shadowed area (remember that the grey-shadowed area corresponds to the parameter space allowed by the tidal deformability of GW170817) \replaced{do not overlap}{are not coincide}, which means that for the standard MIT bag model, SSs are ruled out by the observations \added{concerning} compact stars.

Here we want to stress that, besides the standard MIT bag model, SSs are ruled out by the observations of compact stars for the density-dependent quark mass model \cite{Yang2021b} and the quasi-particle model \cite{Cai2021} if one does not consider alternative theories of gravity or the existence of a DM core in the stars.

\section{SSs in the framework of non-Newtonian gravity}\label{ssnon}

The inverse-square-law of gravity is expected to be modified because of the geometrical effect of the extra space-time dimensions predicted by string theory, which tries to unify gravity with other three fundamental forces \cite{Fischbach1999,Adelberger2003,Adelberger2009}. Non-Newtonian gravity also arises due the exchange of weakly interacting bosons in the super-symmetric extension of SM \cite{Fayet1980,Fayet1981}. 
Many efforts have been tried to constrain the deviations from Newton's gravity, see Ref. \cite{Murata2015} for reviews. 

The effects of non-Newtonian gravity on the properties of compact stars have been widely investigated \cite{Krivoruchenko2009,Wen2009,Wen2011,Sulaksono2011,Zhang2011,Yan2013,Lin14,Lu2017,Yu2018,Yang2020,Yang2021b,Pi2022a}.
The inclusion of non-Newtonian gravity leads to stiffer EOSs, which can support higher maximum masses of compact stars. Thus, some soft EOS of dense nuclear matter cannot be ruled out by the observed massive pulsars considering the non-Newtonian gravity effects \cite{Wen2009}.

\subsection{EOS of SQM in the framework of non-Newtonian gravity}

The deviation from the inverse-square-law of gravity is often characterized by a Yukawa potential \cite{Fujii1971} \footnote{In the weak-field limit, a Yukawa term also appears in alternative theories of gravity such as f(R), the nonsymmetric gravitational theory, and Modified Gravity \cite{Li2019}.}. 
Considering the non-Newtonian gravity effects, the potential energy describing the interaction between the two objects with masses $m_{1}$ and $m_{2}$ is
\begin{equation}
V(r)=-\frac{G m_{1}m_{2}}{r} \left( 1+\alpha e^{-r/\lambda}
\right) = V_N(r) + V_Y(r),
\label{vr}
\end{equation}
where $V_N(r)$ is the Newtonian potential, $V_Y(r)$ is the Yukawa correction,  $G$ is the gravitational constant, $\alpha$ is the dimensionless strength parameter of the Yukawa force, and $\lambda$ is the length scale of the Yukawa force.

In the boson exchange picture, the range of the Yukawa force is 
\begin{equation}
\lambda=\frac{1}{\mu},
\label{mulam}
\end{equation}
where $\mu$ is the mass of the bosons exchanged between $m_1$ and $m_2$.
While, the strength parameter is
\begin{equation}
\alpha=\pm \frac{g^{2}}{4\pi G m_{b}^{2}},
\end{equation}
where the $\pm$ sign refers to scalar/vector bosons \footnote{Scalar bosons lead to a softer EOS of dense matter, while vector bosons make the EOS stiffer \cite{Krivoruchenko2009}. In the following of this review, we will focus on vector bosons. The reason is that a stiff EOS is needed to explain the large mass of PSR J0740+6620.}, $g$ is the boson-baryon coupling constant, and $m_{b}$ is the baryon mass.

\begin{figure}[H]
\includegraphics[width=10.5 cm]{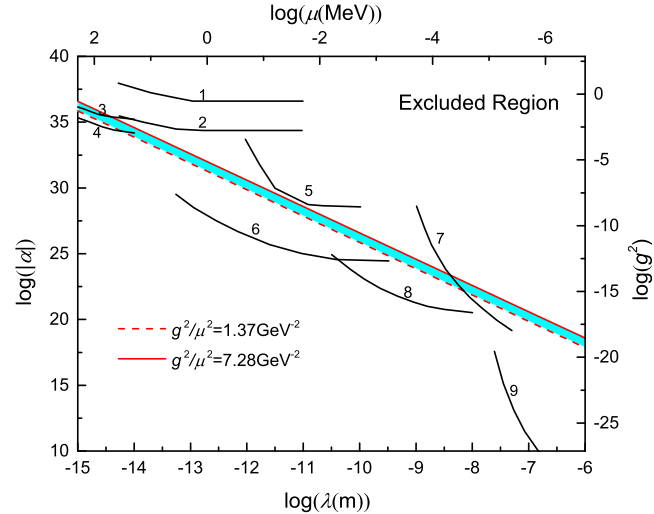}
\caption{Theoretical bounds on $g^{2}/\mu^{2}$ in comparison with constraints on the strength parameter $|\alpha|$ and the range of the Yukawa force $\lambda$ from different experiments: curves 1 and 2 are from Ref. \cite{Kamyshkov2008}, 3 and 4 are from Ref. \cite{Xu2013}, 5 and 6 are from Ref. \cite{Pokotilovski2006}, and 7, 8, 9 are from Refs. \cite{Klimchitskaya2020,Kamiya2015,Chen2016}, respectively. Image from Ref. \cite{Yang2020}. \label{nn_fig}}
\end{figure}   

As will be shown later in Sect.~\ref{allparanon}, SSs could exist for 1.37 GeV$^{-2}\leq g^{2}/\mu^{2}\leq$ 7.28 GeV$^{-2}$. This theoretical region is compared with the constraints from terrestrial experiments in Fig.\ \ref{nn_fig}. One can see that the theoretical region is allowed by many experiments.

A neutral weakly coupled spin-1 gauge U-boson is suggested to be a candidate for the exchanged boson. This U-boson is proposed in the super-symmetric extension of SM \cite{Fayet1980,Fayet1981}, and many terrestrial experiments have been tried to search for it \cite{Yong2013}. It is also found that this U-boson can help to explain the 511 keV $\gamma$-ray observation from the galatic bulge \cite{Jean2003,Boehm2004a,Boehm2004b}.

The extra energy density results from the Yukawa correction $V_Y(r)$ of Eq.\ (\ref{vr}) is \cite{Long2003,Lu2017,Yang2020}
\begin{equation}
\epsilon_{Y}=\frac{1}{2V}\int 3n_{b}(\vec{x}_{1})\frac{g^{2}}{4\pi}
\frac{e^{-\mu r}}{r}3n_{b}(\vec{x}_{2})d\vec{x}_{1}d\vec{x}_{2} = \frac{9}{2} \frac{g^{2}}{\mu^{2}}n_{b}^{2} ,
\label{inte1}
\end{equation}
where $n_b(\vec x_1)$ and $n_b(\vec x_2)$ are the densities, $r = |\vec{x}_{1} - \vec{x}_{2}|$, and $V$ is the normalization volume. The prefactor of 3 appears before $n_b(\vec x_1)$ and $n_b(\vec x_2)$ because the baryon number of quarks is $1/3$. 
The extra pressure results from the Yukawa correction is
\begin{equation}
p_{Y}=n_{b}^{2}\frac{d}{dn_{b}}\bigg(\frac{\epsilon_{Y}}{n_{b}}\bigg).
\end{equation}
Assuming that the boson mass is independent of the density, one gets
\begin{equation}
p_{Y}=\epsilon_{Y} = \frac{9}{2} \frac{g^{2}}{\mu^{2}}n_{b}^{2}.
\end{equation}
Thus, the total energy density and pressure of SQM are 
\begin{equation}
\epsilon = \epsilon_{Q} + \epsilon_{Y},
\end{equation}
\begin{equation}
p = p_{Q} + p_{Y},
\end{equation}
where $\epsilon_{Q}$ and $p_{Q}$ are given by Eqs.\ (\ref{eq:epsQ}) and (\ref{eq:pQ}), respectively.

When we emlpoy the EOS of SQM described by $p(\epsilon)$, the Yukawa correction is considered as a part of the matter system in GR
\begin{equation}
T^{\alpha\beta} = \left[
\epsilon + p(\epsilon) \right] u^\alpha u^\beta + p(\epsilon) g^{\alpha\beta}.
\end{equation} 
As a result, the effects of non-Newtonian gravity on compact stars can be studied by solving the TOV equations \cite{Fujii1988,Krivoruchenko2009,Wen2009}.

\subsection{The allowed parameter space of SQM in the framework of non-Newtonian gravity} \label{allparanon}

\begin{figure}[H]
\includegraphics[width=10.5 cm]{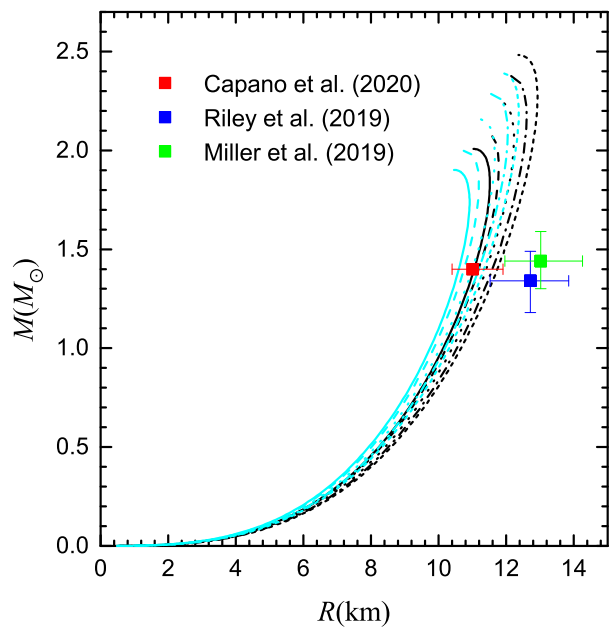}
\caption{The mass-radius relation of SSs for $m_{s}=95$ MeV. The black and cyan lines are for parameter sets of [$B^{1/4}$(MeV), $\alpha_{S}$] with (142, 0.2) and (146, 0), respectively. For each color, the lines are for $g^{2}/\mu^{2}=0$, 1, 3, 5, and 7 GeV$^{-2}$ from left to right. The observational data are the same as those shown in Fig.\ \ref{rm} for GW171807 and PSR J0030+0451. Image from Ref. \cite{Yang2020}. \label{fig1}}
\end{figure}

\begin{figure}[H]
\begin{adjustwidth}{-\extralength}{0cm}
\centering
\includegraphics[width=13.5cm]{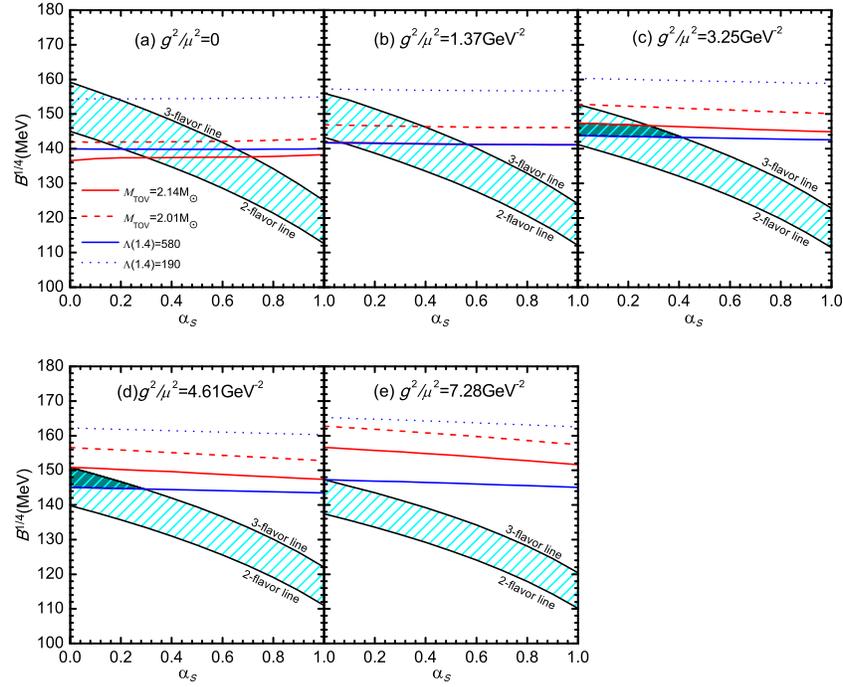}
\end{adjustwidth}
\caption{Constraints on the parameters of the standard MIT model for $m_{s}=95$ MeV and different values of $g^{2}/\mu^{2}$. Image from Ref. \cite{Yang2020}. 
\label{fig2}}
\end{figure}

Considering the non-Newtonian gravity effects, the mass-radius relation of SSs is shown in Fig.\ \ref{fig1}. Both parameter sets of [$B^{1/4}$(MeV), $\alpha_{S}$] used in Fig.\ \ref{fig1} can satisfy the "2-flavor line" and "3-flavor line" constraint as will be shown in Fig.\ \ref{fig2}. One can find that the inclusion of the non-Newtonian gravity leads to a larger maximum mass of SSs.

By imposing the first four constraints presented in Sect.~\ref{constnon} (i.e., the last constraint from the observation of PSR J0030+0451 is not considered here), the allowed parameter space of the standard MIT model is restricted to the dark cyan-shadowed regions shown in Figs.\ \ref{fig2}(c), \ref{fig2}(d), which correspond to $g^{2}/\mu^{2}=3.25$ GeV$^{-2}$ and 4.61 GeV$^{-2}$, respectively. 

As can be seen from Fig.\ \ref{fig2}(a), the constraints $M_{\rm max}\geq 2.14\, M_{\odot}$ and $\Lambda(1.4)\leq 580$ cannot be satisfied simultaneously for the case of $g^{2}/\mu^{2}=0$. However, the gap between the $M_{\rm max}=2.14\, M_{\odot}$ line and the $\Lambda(1.4)=580$ line becomes smaller as the value of $g^{2}/\mu^{2}$ increases, and finally these two lines almost completely \replaced{overlap with each other}{coincide} when $g^{2}/\mu^{2}$ is as large as 1.37 GeV$^{-2}$ [see Fig.\ \ref{fig2}(b)], which means all the four constraints can be satisfied. The allowed parameter space continues to exist as the value of $g^{2}/\mu^{2}$ increases until it is vanished for $g^{2}/\mu^{2} > 7.28$ GeV$^{-2}$[see Fig.\ \ref{fig2}(e)], where the "2-flavor line" and "3-flavor line" constraint cannot be satisfied simultaneously. Therefore, we have the conclusion that SSs can exist for 1.37 GeV$^{-2}\leq g^{2}/\mu^{2}\leq$ 7.28 GeV$^{-2}$.

Moreover, one can find that for the existence of SSs, $B^{1/4}$ must be larger than 141.3 MeV, and $\alpha_{S}$ must be smaller than 0.56 [see Fig.\ \ref{fig2}(b), where the $M_{\rm max}=2.14\, M_{\odot}$ line (the $\Lambda(1.4)=580$ line) meets the 3-flavor line at the point (141.3, 0.56)], while the upper limit of $B^{1/4}$ is 150.9 MeV [see Fig.\ \ref{fig2}(d), where the $M_{\rm max}=2.14\, M_{\odot}$ line cuts across the 3-flavor line at the point (150.9, 0)] \cite{Yang2020}.

\added{We also find that the largest allowed maximum mass of SSs is $2.37\, M_{\odot}$, corresponding to the parameter set $g^{2}/\mu^{2}=7.28$ GeV$^{-2}$, $\alpha_{S}=0$ and $B^{1/4}=147.3$ MeV. Thus, the GW190814’s secondary component with mass $2.59_{-0.09}^{+0.08}\, M_{\odot}$ \cite{Abbott2020} could not be a static SS even considering the non-Newtonian effect. However, it could be a rotating SS \cite{Zhou2019}.}

\begin{figure}[H]
\begin{adjustwidth}{-\extralength}{0cm}
\centering
\includegraphics[width=13.5cm]{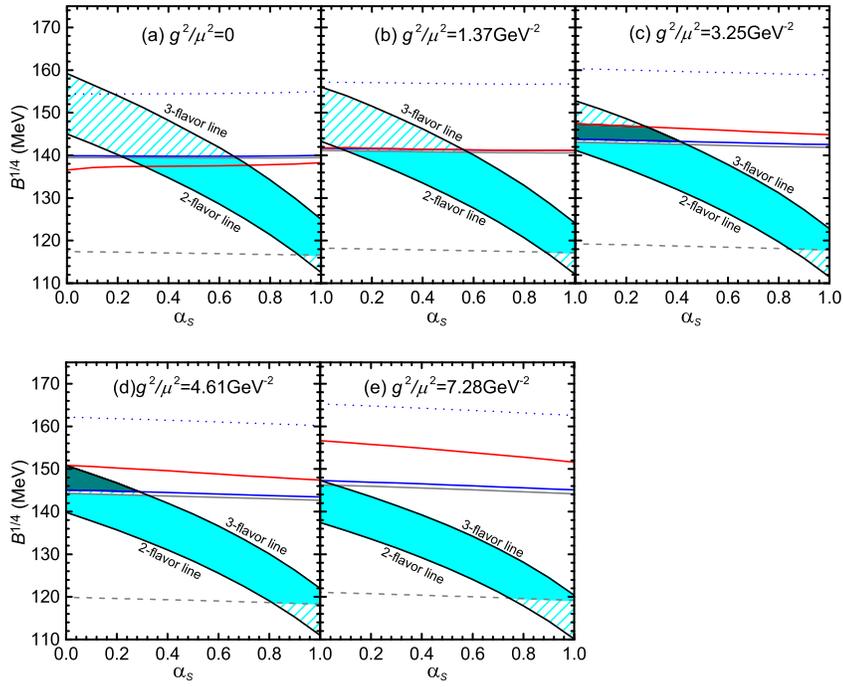}
\end{adjustwidth}
\caption{Constraints on the parameters of the standard MIT model for $m_{s}=95$ MeV. The red and blue lines are the same as those in Fig.\ \ref{fig2} except that the dashed red line is not shown in this figure. The dark cyan-shadowed regions are also the same as that in Fig.\ \ref{fig2}, which are restricted by the first four constraints presented in Sect.~\ref{constnon}. The last constraint (i.e., the constraint from the observation of PSR J0030+0451) leads to the cyan-shaded region. This figure is for the case of Riley et al. \cite{Riley2019}, where the grey solid and dashed lines are for the [$M\, (M_{\odot}$), $R$\,(km)] sets (1.49, 11.52) and (1.18, 13.85), respectively. Image from Ref. \cite{Pi2022a}. \label{fig_Riley}}
\end{figure}

\begin{figure}[H]
\begin{adjustwidth}{-\extralength}{0cm}
\centering
\includegraphics[width=13.5cm]{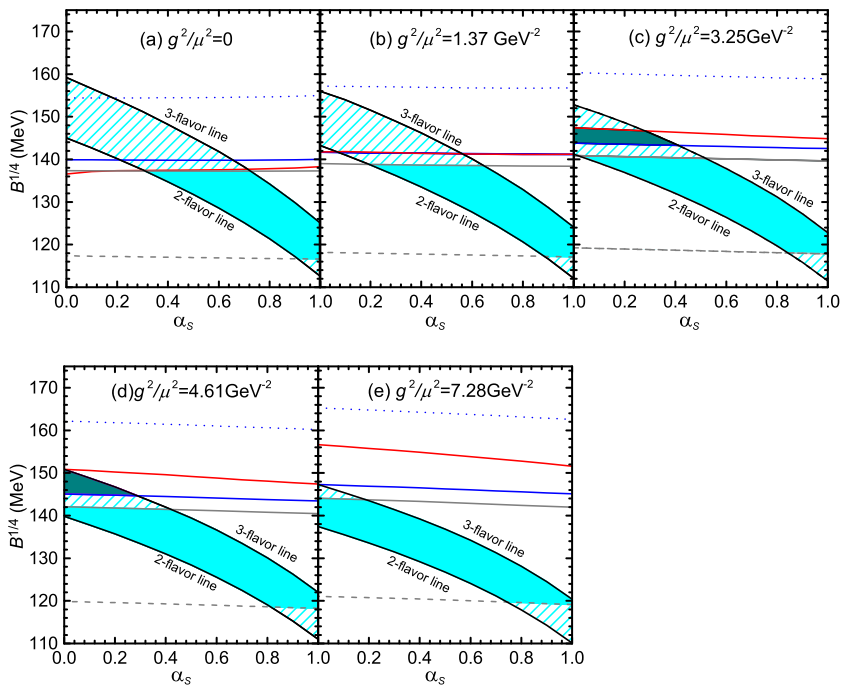}
\end{adjustwidth}
\caption{The same as Fig.\ \ref{fig_Riley}, but for the case of Miller et al.\ \cite{Miller2019}, where the grey solid and dashed lines are for the [$M\, (M_{\odot}$), $R$\,(km)] sets (1.59, 11.96) and (1.30, 14.26), respectively. Image from Ref. \cite{Pi2022a}. \label{fig_Miller}}
\end{figure}

Now, we consider all the five constraints presented in Sect.~\ref{constnon}. From Fig.\ \ref{fig_Riley} and Fig.\ \ref{fig_Miller}, one sees that the gap between the blue solid line and the grey solid line is almost unchanged as the value of $g^{2}/\mu^{2}$ increases. As a result, the dark cyan-shadowed regions and the cyan-shadowed regions cannot overlap for all the choices of the value of $g^{2}/\mu^{2}$, which means that for the standard MIT model, SSs are ruled out by the observations of compact stars even if the effects of non-Newtonian gravity are considered.

\section{SSs with a mirror-dark-matter core}\label{ssmdm}

As a consequence of the parity symmetric extension of SM of particles, mirror dark matter (MDM) is regarded as a kind of DM candidate \cite{Foot1991}. 
The idea of MDM was first realized by Lee and Yang in 1956 when the weak interaction was found to violate parity \cite{Lee1956}. These authors suggested that the existence of a set of unknown particles could restore the symmetry. There are many reviews about MDM \cite{Foot2004,Berezhiani2004,Berezhiani2005,Okun2007,Foot2014}. 

The properties of MDM admixed compact stars have been widely investigated \cite{Sandin2009,Ciarcelluti2011,Berezhiani2021,Kain2021,Ciancarella2021,Yang2021a,Emma2022,Hippert2022}. It is found that the existence of a MDM core in compact stars leads to a softer EOS. What's more, it is found that the mass and radius observations of compact stars might serve as a signature of the existence of a MDM core in NSs \cite{Ciarcelluti2011}.

\subsection{EOS of MDM}

In the minimal parity-symmetric extension of SM \cite{Foot1991,Pavsic1974,Sandin2009,Ciarcelluti2011}, the ordinary matter and MDM are described by the same lagrangians, and the only difference between them is that ordinary particles have left-handed interactions while mirror particles have right-handed interactions. Therefore, the microphysics of MDM and ordinary matter are exactly the same, which means that they have the same EOS. In this review, MDM is the mirror strange quark matter made of mirror up ($u'$), mirror down ($d'$) and mirror strange ($s'$) quarks and mirror electrons ($e'$), which has the same EOS as that of SQM.

MDM and ordinary matter could interact directly. For example, photon-mirror photon kinetic mixing has been studied and its strength is of order $10^{-9}$ \cite{Foot2014,Ciancarella2021}. This interaction is too weak to have apparent effect on the structure of compact stars \cite{Sandin2009}. The interactions between quarks and mirror quarks have not been studied so far. While, it is reasonable to suppose that these interactions are weak and their effects \replaced{on}{to} the structure of SSs could be ignored \footnote{In the study of neutron-mirror neutron ($n-n'$) mixing \cite{Berezhiani2006,Berezhiani2009,Goldman2019,McKeen2021,Goldman2022}, it is found that a mirror-matter core could develop in ordinary NSs through the process of $n-n'$ conversion \cite{Berezhiani2021}. However, since SQM is self-bound, the transformation to mirror matter is suppressed \cite{Berezhiani2021}.}.

\subsection{The properties of SSs with a MDM core}

For SSs with a MDM core, although the SQM and MDM components do not interact directly, they can interact with each other through the gravitational interaction. Thus, a two-fluid formalism is employed to study the properties of SSs with a MDM core.

The TOV equations in the two-fluid formalism are given by \cite[e.g.,][]{Das2022a,Ciancarella2021,Sandin2009,Ciarcelluti2011}
\begin{equation}
\frac{dm(r)}{dr}=4\pi \epsilon(r) r^{2},
\label{tov1}
\end{equation}
\begin{equation}
\frac{dp_{Q}(r)}{dr}=-\frac{[m(r)+4\pi r^{3}p(r)][\epsilon_{Q}(r)+p_{Q}(r)]}{r[r-2m(r)]},
\label{tov2}
\end{equation}
\begin{equation}
\frac{dp_{M}(r)}{dr}=-\frac{[m(r)+4\pi r^{3}p(r)][\epsilon_{M}(r)+p_{M}(r)]}{r[r-2m(r)]},
\label{tov3}
\end{equation}
with
\begin{eqnarray}
\label{newp}
\epsilon(r) &=& \epsilon_{Q}(r)+\epsilon_{M}(r),\\
p(r) &=& p_{Q}(r)+p_{M}(r),
\label{neweps}
\end{eqnarray}
where the subscript $Q$ and $M$ are for SQM and MDM, respectively.

In the two-fluid formalism, the dimensionless tidal deformability ($\Lambda$) can be calculated similar to that discribed in Sect.~\ref{tid}, except that Eq.\ (\ref{Qr_w}) should be changed into \cite[e.g.,][]{Das2022a,Ciancarella2021}
\begin{eqnarray}
\nonumber Q(r)&=&\frac{4\pi r}{r-2m(r)}
       \bigg[[5-y(r)]\epsilon(r)+[9+y(r)]p(r)
	+\frac{\epsilon_{Q}(r)+p_{Q}(r)}{\partial p_{Q}(r)/\partial \epsilon_{Q}(r)}+\frac{\epsilon_{M}(r)+p_{M}(r)}{\partial p_{M}(r)/\partial \epsilon_{M}(r)} \bigg]
    \\ &&-4\bigg[\frac{m(r)+4\pi r^{3}p(r)}{r[r-2m(r)]}\bigg]^{2}.
\label{Qr}
\end{eqnarray}

To get $y(r)$, Eq.\ (\ref{yr_w}) should be calculated together with Eqs.\ (\ref{tov1}),\ (\ref{tov2}),\ (\ref{tov3}) for given SQM and MDM pressure at the center of the star, and the boundary conditions are: $y(0)=2$, $m(0)=0$, $p_{Q}(R)=0$, $p_{M}(R_M)=0$ ($R$ is the radius of SSs, $R_M$ is the radius of the MDM core).

Besides the energy density jump at the surface of SS, another energy density jump $\epsilon_{sM}$ exists at the surface of the MDM core. As a result, an additional term $-4\pi R_M^{3}\epsilon_{sM}/M(R_M)$ should be added to $y(R_M)$.

Obviously, the properties of SSs with a MDM core change with the mass fraction of MDM ($f_{M}$), where $f_{M} \equiv M_{M}/M$ ($M$ is the total mass of the star, and $M_{M}$ is the mass of the MDM core).

\begin{figure}[H]
\includegraphics[width=10.5 cm]{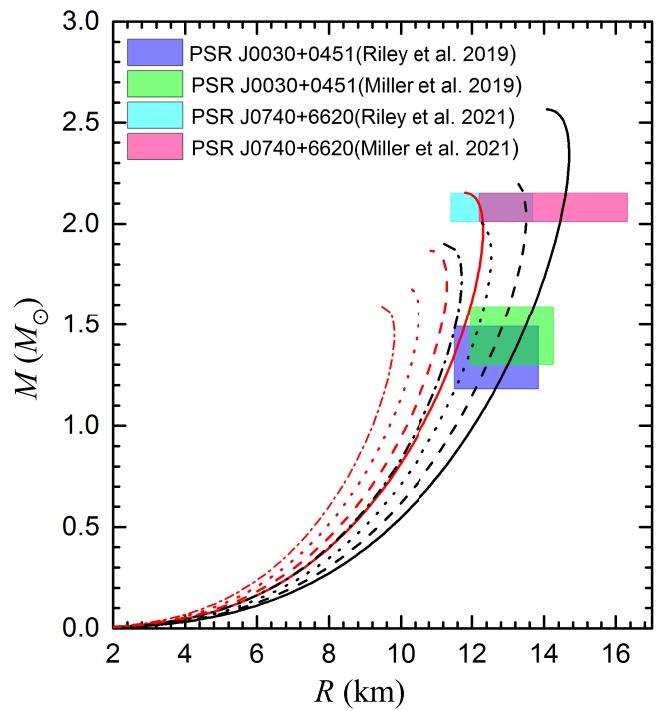}
\caption{The mass-radius relation of SSs for $m_{s}=93$ MeV and $\alpha_{S}=0.7$. The black and red lines are for $B^{1/4}=125.1$ MeV and $137.3$ MeV, respectively. For each color, the lines are for $f_{M}=0$, 10\%, 20\%, and 30\% from right to left. The observational data are the same as those shown in Fig.\ \ref{rm} for PSR J0030+0451 and PSR J0740+6620. Image from Ref. \cite{Yang2021a}.\label{MDM_fig1}}
\end{figure}   

Fig.\ \ref{MDM_fig1} shows the mass-radius relation of SSs for different values of $f_{M}$. Two sets of SQM parameters [$B^{1/4}$(MeV), $\alpha_{S}$] [(125.1, 0.7) and (137.3, 0.7)] are choosen, which are shown by the magenta dots in Fig.\ \ref{constraints}(b). From Fig.\ \ref{MDM_fig1}, one sees that for fixed values of $B^{1/4}$ and $\alpha_{S}$, the maximum mass of SSs decreases as the value of $f_{M}$ increases. In other words, the existence of a MDM core leads to a softer EOS.

\begin{figure}[H]
\includegraphics[width=10.5 cm]{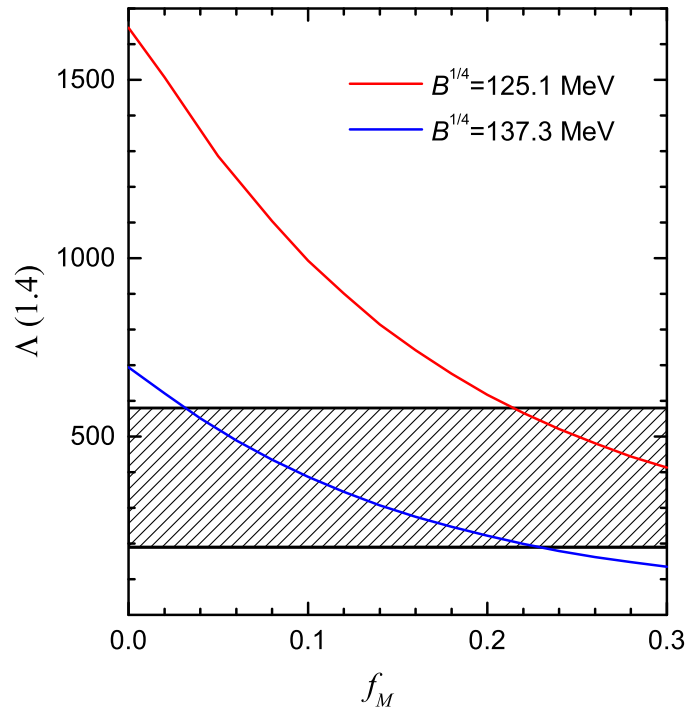}
\caption{Relation between $\Lambda(1.4)$ and $f_{M}$ for $m_{s}=93$ MeV. The shaded region corresponds to the observation of GW170817 [$70<\Lambda(1.4)<580$]. Image from Ref. \cite{Yang2021a}. \label{MDM_fig2}}
\end{figure}

Fig.\ \ref{MDM_fig2} shows the relation between $\Lambda(1.4)$ and $f_{M}$. One sees that for a given value of $f_{M}$, the value of $\Lambda(1.4)$ increases with the decreasing of $B^{1/4}$. It also can be seen from Fig.\ \ref{MDM_fig2} that $\Lambda(1.4)$ decreases as the value of $f_{M}$ increases. For both parameter sets considered in Fig.\ \ref{MDM_fig2}, SSs cannot agree with the observation of GW170817 if $f_{M}$ is not large enough. The critical values of $f_{M}$ are 3.1\% and 21.4\% for $B^{1/4}=137.3$ MeV and $125.1$ MeV, respectively.

\subsection{The allowed parameter space of SQM for SSs with a MDM core}

\begin{figure}[H]
\begin{adjustwidth}{-\extralength}{0cm}
\centering
\includegraphics[width=13.5cm]{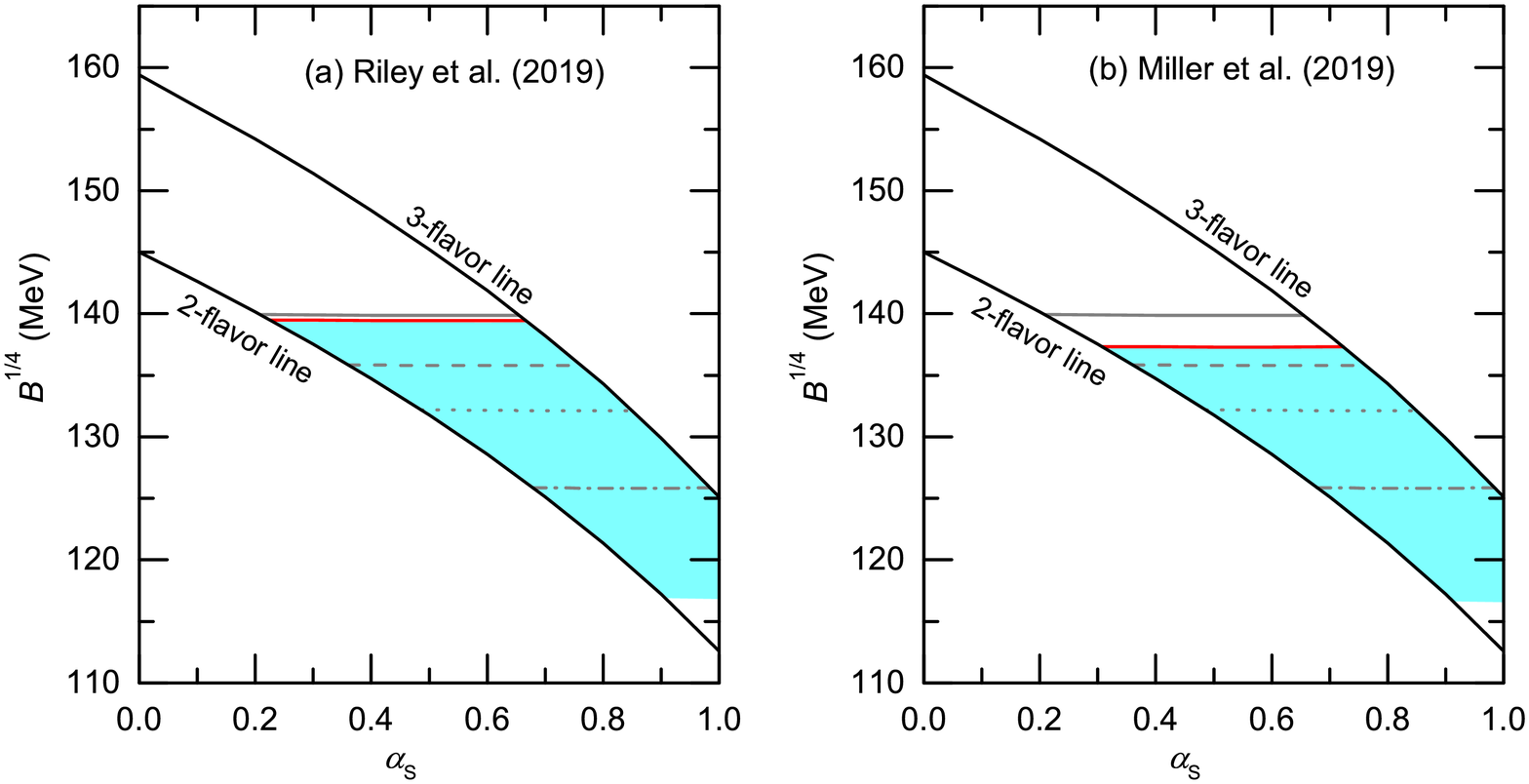}
\end{adjustwidth}
\caption{Constraints on the parameters of the standard MIT model for $m_{s}=93$ MeV. The cyan-shadowed areas are the same as those in Fig.\ \ref{constraints}. The grey lines are for $\Lambda(1.4)=580$ with $f_{M}=0$, 5\%, 10\%, and 20\% from top to bottom. The red lines also correspond to $\Lambda(1.4)=580$, which are for $f_{M}=0.5$\% in (a), and for $f_{M}=3.1$\% in (b). Image from Ref. \cite{Yang2021a}. \label{MDM_fig4}}
\end{figure}

The allowed parameter space of the standard MIT model is investigated by imposing all the five constraints presented in Sect.~\ref{constnon}, and the result is shown in Fig.\ \ref{MDM_fig4}. \replaced{Note}{Mention} that the cyan-shadowed areas in Fig.\ \ref{MDM_fig4} are for the case of SSs without a MDM core, which could fulfill the constraints from the observaions of PSR J0740+6620 and PSR J0030+0451 (see Sect.~\ref{constnon}). 
From Fig.\ \ref{MDM_fig4}, one sees that the parameter space area which satisfies the tidal deformability observation of GW170817 [the area above the $\Lambda(1.4)=580$ line] shifts downward as the value of $f_{M}$ increases, and that area \replaced{begins}{begin} to overlap with the cyan-shadowed area for $f_{M}=0.5$\% for the case of Riley et al.\ \cite{Riley2019} [Fig.\ \ref{MDM_fig4}(a)], and $f_{M}=3.1$\% for the case of Miller et al.\ \cite{Miller2019} [Fig.\ \ref{MDM_fig4}(b)]. 

Thus, we have the conclusion that assuming PSR J0740+6620 and PSR J0030+0451 do not have a MDM core, all the observations of compact stars could be satisfied if SSs in GW170817 have a large enough MDM core ($f_{M}>0.5$\% for Riley et al. \cite{Riley2019}, and $f_{M}>3.1$\% for Miller et al. \cite{Miller2019}). However, PSR J0740+6620 or PSR J0030+0451 might also have a MDM core. In that case, one can easily decuce that SSs in GW170817 should have a larger MDM core in order to satisfy all the observations of compact stars. 

\section{Conclusions}\label{conclusions}

In this review, it is shown that for the standard MIT bag model, SSs are ruled out by the observations from GW170817, PSR J0740+6620 (and PSR J0030+0451). In fact,  SSs are also ruled out for the density-dependent quark mass model \cite{Yang2021b} and the quasi-particle model \cite{Cai2021}. However, the tension between the theory of SSs and the observations of compact stars could be resolved if alternative gravity (e.g. non-Newtonian gravity) or dark matter (e.g. mirror dark matter) is considered.  

We find that non-Newtonian gravity effects of SSs could help to relieve the tension between the observations of the tidal deformability of GW170817 and the mass of PSR J0740+6620. However, if the constraints from the mass and radius of PSR J0030+0451 is added, the existence of SSs is still ruled out even if the effects of non-Newtonian gravity is considered. However, the possibility of the existence of SSs should be investigated under the framework of other alternative gravity theories \cite{Li2019,Shao2019,Olmo2020}.

In the scenario of SSs with a MDM core, it is found that to explain all the observations of compact stars, a MDM core shoud exist in SSs of GW170817. Moreover, although this result is derived for the case of MDM, it is qualitatively valid for other kinds of DM that could exist inside SSs. As a result, one has the conclusion that for the standard MIT bag model, the current observations of compact stars could serve as an evidence for the existence of a DM core inside SSs.

\begin{figure}[H]
\includegraphics[width=10.5 cm]{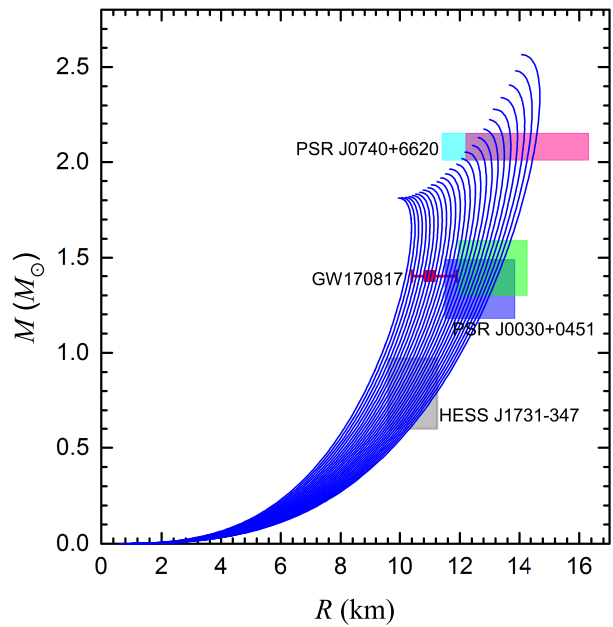}
\caption{The mass-radius relation of SSs for $m_{s}=93$ MeV, $\alpha_{S}=0.7$ and $B^{1/4}=125.1$ MeV. The rightmost line is for SSs without a MDM core (i.e., $f_{M}=0$). Other lines are for SSs with different $f_{M}$, namely, from $f_{M}=2$\% to 50\% with step 2\% (from right to left. For $f_{M}=50$\%, the MDM core has a same radius as the SS with ordinary SQM). The observational data are the same as those shown in Fig.\ \ref{rm}.\label{rm_MDM}}
\end{figure}   

One obvious difference between the non-Newtonian scenario and the MDM scenario is that the equilibrium sequence of SSs with non-Newtonian gravity effects is a single line, while, the equilibrium sequence of SSs with a MDM core is non-unique, which is a fan-shaped region as shown in Fig.\ \ref{rm_MDM}. Since the value of $f_{M}$ for each SS is different (which depends on its evolutionary history) \cite{Ciarcelluti2011}, SSs could be located everywhere on that region. With more mass and radius observations of compact stars in the near future, one can distinguish these two scenarios.

\added{From the aspect of the study of SQM, this review is limited to the standard MIT bag model, the density-dependent quark mass model \cite{Backes2021,Chakrabarty1989,Benvenuto1995,Lugones1995,Peng2000,Xia2014}, and the quasi-particle model \cite{Peshier1994,Gorenstein1995,Peshier2000,Zong2008,Li2019c}. However, similar investigations should be carried out in the future, which are based on other phenomenological SQM models such as the Nambu-Jona-Lasinio (NJL) model \cite{Buballa2005,Ruster2004,Menezes2006,Klahn2013,Chu2016}, the perturbative QCD approach \cite{Freedman1977,Freedman1978,Kurkela2010} and the models considering isospin interaction \cite{Chu2014,Liu2016,Chen2017,Chu2019}. Moreover, SQM is supposed to be in a color superconducting state \cite{Alford1998,Rapp1998,Rapp2000,Rajagopal2001b,Alford2001c,Alford2008}, the quarks might pair in different patterns such as CFL (color-flavor locked phase) \cite{Alford1999,Rajagopal2001,Lugones2002} and 2SC (two-flavor superconducting) \cite{Alford1998,Rapp1998}. Therefore, simlliar investigations should also be carried out for color superconducting SQM. Note that recently SSs made of CFL SQM were employed to explain the large mass of the GW190814’s secondary component ($2.59_{-0.09}^{+0.08}\, M_{\odot}$) \cite{Bombaci2021,Roupas2021,Horvath2021,Miao2021c}.}  

In summary, alternative gravity theory and DM could play a significant role in the study of SSs and the possibility of the existence of SSs could provide a clue to new physics. More observations of compact stars are expected in the near future by the gravational wave detectors aLIGO, aVirgo, Kagra, the Einstein Telescope (ET) and the Cosmic Explorer (CE); the X-ray missions eXTP and STROBE-X; and the radio observatory SKA. With these new data, we might finally confirm (or rule out) the existence of SSs \footnote{\added{In this review, we mainly discuss SSs by considering the bulk properties of compact stars such as the mass, radius, and tidal deformability. However, the study of SSs is far beyond these issues. For example, some special phenomena associated with compact stars might also help to reveal their nature. In fact, the neutrinos emitted during the combustion of an NS into an SS could be possibly directly detected \cite{Pagliara2013}; some explosions, especially fast radio bursts, may point to the existence of SSs \cite{Geng2021}; and it is also suggested that people could try to identify strange quark objects by searching for close-in exoplanets around pulsars \cite{Kuerban2020}.}} and provide useful informations about the gravity theory or the properties of DM.

\vspace{6pt} 




\funding{This work is supported by the National Key R\&D Program of China under Grant No. 2021YFA0718504, the National SKA Program of China under Grant No. 2020SKA0120300, and the Scientific Research Program of the National Natural Science Foundation of China under Grant No. 12033001. F.W. acknowledges support from the U.S. National Science Foundation under Grant PHY-2012152.}

\institutionalreview{Not applicable.}

\informedconsent{Not applicable.}

\dataavailability{Observational data used in this paper are quoted from the cited works.} 

\acknowledgments{\added{The authors are grateful to the anonymous referees for their useful suggestions greatly helped us to improve the paper.}}

\conflictsofinterest{The authors declare no conflict of interest.}


\abbreviations{Abbreviations}{
The following abbreviations are used in this review:\\
\noindent 
\begin{tabular}{@{}ll}
DM & Dark mtter \\
EOS & Equation of state \\
eXTP & enhanced X-ray Timing and Polarimetry mission\\
GR & General Relativity \\
LIGO & Laser interferometer gravitational wave observatory\\
MDM & Mirror dark matter \\
MIT & Massachusetts Institute of Technology \\
NICER & Neutron star Interior Composition Explorer \\
NS & Neutron star \\
SM & Standard Model \\
SQM & Strange quark matter \\
SS & Strange star \\
STROBE-X & Spectroscopic Time-Resolving Observatory for Broadband Energy X-rays \\
SKA & Square Kilometre Array observatory \\
TOV & Tolman-Oppenheimer-Volkoff \\
Virgo & Virgo interferometer 


\end{tabular}
}

\begin{adjustwidth}{-\extralength}{0cm}

\reftitle{References}

\end{adjustwidth}

\begin{thebibliography}{999}

\bibitem[]{Mann2020}
Mann, A. The golden age of neutron-star physics has arrived. {\em Nature} {\bf 2020}, {\em 579}, 20--22.

\bibitem[]{Ozel2016}
\"Ozel, F.; Freire, P. Masses, radii, and the equation of state of neutron stars. {\em Annu. Rev. Astron. Astrophys.} {\bf 2016}, {\em 88}, 401--440.
\bibitem[]{Lattimer2016}
Lattimer, J.M.; Prakash, M. The equation of state of hot, dense matter and neutron stars. {\em Phys. Rep.} {\bf 2016}, {\em 621}, 127--164.
\bibitem[]{Lattimer2019}
Lattimer, J.M. Neutron Star Mass and Radius Measurements. {\em Universe} {\bf 2019}, {\em 5}, 159.
\bibitem[]{Raithel2019}
Raithel, C.A. Constraints on the Neutron Star Equation of State from GW170817, {\em Eur. Phys. J. A} {\bf 2019}, {\em 55}, 80.
\bibitem[]{Li2019}
Li, B.-A.; Krastev, P.G.; Wen, D.-H.; Zhang, N.-B. Towards understanding astrophysical effects of nuclear symmetry energy. {\em Eur. Phys. J. A} {\bf 2019}, {\em 55}, 117.
\bibitem[]{Orsaria2019}
Orsaria, M.G.; Malfatti, G.; Mariani, M.; Ranea-Sandoval, I.F.; Garc\'a, F.; Spinella, W.M.; Contrera, G.A.; Lugones, G.; Weber, F. Phase transitions in neutron stars and their links to gravitational waves. {\em J. Phys. G: Nucl. Part. Phys.} {\bf 2019}, {\em 46}, 073002.
\bibitem[]{Baiotti2019}
Baiotti, L. Gravitational waves from neutron star mergers and their relation to the nuclear equation of state. {\em Prog. Part. Nucl. Phys.} {\bf 2019}, {\em 109}, 103714.
\bibitem[]{Li2020}
\added{Li, A.; Zhu, Z.-Y.; Zhou, E.-P.; Dong, J.-M.; Hu, J.-N.; Xia, C.-J. Neutron star equation of state: Quark mean-field (QMF) modeling and applications. {\em J. High Energy Astrophys.} {\bf 2020}, {\em 28}, 19.}
\bibitem[]{Burgio2020}
Burgio, G.F.; Vida\~na, I. The Equation of State of Nuclear Matter: From Finite Nuclei to Neutron Stars. {\em Universe} {\bf 2020}, {\em 6}, 119.
\bibitem[]{Chatziioannou2020}
Chatziioannou, K. Neutron-star tidal deformability and equation-of-state constraints. {\em Gen. Rel. Grav.} {\bf 2020}, {\em 52}, 109.
\bibitem[]{Lattimer2021}
Lattimer, J.M. Neutron Stars and the Nuclear Matter Equation of State. {\em Annu. Rev. Nucl. Part. Sci.} {\bf 2021}, {\em 71}, 433--464.
\bibitem[]{Li2021}
Li, B.-A.; Cai, B.-J.; Xie, W.-J.; Zhang, N.-B. Progress in Constraining Nuclear Symmetry Energy Using Neutron Star Observables Since GW170817. {\em Universe} {\bf 2021}, {\em 7}, 182.




\bibitem[]{Farhi1984}
Farhi, E.; Jaffe, R.L. Strange matter. {\em Phys. Rev. D} {\bf 1984}, {\em 30}, 2379--2390.
\bibitem[]{Haensel1986}
Haensel, P.; Zdunik, J.L.; Schaefer, R. Strange quark stars. {\em Astron. Astrophy.} {\bf 1986}, {\em 160}, 121--128.
\bibitem[]{Alcock1986}
Alcock, C.; Farhi, E.; Olinto, A. Strange Stars. {\em Astrophys. J.} {\bf 1986}, {\em 310}, 261.
\bibitem[]{Alcock1988}
Alcock, C.; Olinto, A. Exotic phases of hadronic matter and their astrophysical application. {\em Annu. Rev. Nucl. Part. Sci.} {\bf 1988}, {\em 38}, 161--184.
\bibitem[]{Madsen1999}
Madsen, J. Physics and Astrophysics of Strange Quark Matter. {\em Lect. Notes Phys.} {\bf 1999}, {\em 516}, 162--203.
\bibitem[]{Weber2005}
Weber, F. Strange quark matter and compact stars. {\em Prog. Part. Nucl. Phys.} {\bf 2005}, {\em 54}, 193--288.
\bibitem[]{Itoh1970}
Itoh, N. Hydrostatic Equilibrium of Hypothetical Quark Stars. {\em Prog. Theor. Phys.} {\bf 1970}, {\em 44}, 291--292.
\bibitem[]{Bodmer1971}
Bodmer, A.R. Collapsed Nuclei. {\em Phys. Rev. D} {\bf 1971}, {\em 4}, 1601--1606.
\bibitem[]{Witten1984}
Witten, E. Cosmic separation of phases. {\em Phys. Rev. D} {\bf 1984}, {\em 30}, 272--285.
\bibitem[]{Terazawa1989a}
Terazawa, H. Super-Hypernuclei in the Quark-Shell Model. {\em J. Phys. Soc. Jpn.} {\bf 1989}, {\em 58}, 3555--3563.
\bibitem[]{Terazawa1989b}
Terazawa, H. Super-Hypernuclei in the Quark-Shell Model. II. {\em J. Phys. Soc. Jpn.} {\bf 1989}, {\em 58}, 4388--4393.
\bibitem[]{Glendenning1990}
Glendenning, N.K. Fast Pulsars, Strange Stars:. An Opportunity in Radio Astronomy. {\em Mod. Phys. Lett. A} {\bf 1990}, {\em 5}, 2197--2207.
\bibitem[]{Caldwell1991}
Caldwell, R.R.; Friedman, J.L. Evidence against a strange ground state for baryons. {\em Phys. Lett. B} {\bf 1991}, {\em 264}, 143--148.  
\bibitem[]{Bhattacharyya2016}
Bhattacharyya, S.; Bombaci, I.; Logoteta, D.; Thampan, A.V. Fast spinning strange stars: possible ways to constrain interacting quark matter parameters. {\em Mon. Not. R. Astron. Soc.} {\bf 2016}, {\em 457}, 3101--3114.
\bibitem[]{Di Clemente2022}
Di Clemente, F.; Drago, A.; Pagliara, G. Is the compact object associated with HESS J1731-347 a strange quark star? arXiv 2022, arXiv:2211.07485.
\bibitem[]{Horvath2023}
\added{Horvath, J.E.; Rocha, L.S.; de S\'a, L.M.; Moraes, P.H.R.S.; Bar\~ao, L.G.; de Avellar, M.G.B.; Bernardo, A.; Bachega, R.R.A. A light strange star in the remnant HESS J1731-347: minimal consistency checks. arXiv 2023, arXiv:2303.10264.}

\bibitem[]{Doroshenko2022}
Doroshenko, V.; Suleimanov, V.; P\"uhlhofer, G.; Santangelo, A. A strangely light neutron star within a supernova remnant. {\em Nat. Astron.} {\bf 2022}, {\em 6}, 1444--1451.
\bibitem[]{Suwa2018}
Suwa, Y.; Yoshida, T.; Shibata, M.; Umeda, H.; Takahashi, K. On the minimum mass of neutron stars.  {\em Mon. Not. Roy. Astron. Soc.} {\bf 2018}, {\em 481}, 3305--3312. 
\bibitem[]{Li1999a}
\added{Li, X.-D.; Bombaci, I.; Dey, M.; Dey, J.; van den Heuvel, E.P.J. Is SAX J1808.4-3658 a Strange Star? {\em Phys. Rev. Lett.} {\bf 1999}, {\em 83}, 3776--3779.}
\bibitem[]{Li1999b}
\added{Li, X.-D.; Ray, S.; Dey, J.; Dey, M.; Bombaci, I. On the Nature of the Compact Star in 4U 1728-34. {\em Astrophys. J.} {\bf 1999}, {\em 527}, L51--L54.}
\bibitem[]{Drake2002}
\added{Drake, J.J.; Marshall, H.L.; Dreizler, S.; Freeman, P.E.; Fruscione, A.; Juda, M.; Kashyap, V.; Nicastro, F.; Pease, D.O.; Wargelin, B.J.; Werner, K. Is RX J1856.5-3754 a Quark Star? {\em Astrophys. J.} {\bf 2002}, {\em 572}, 996--1001.}
\bibitem[]{Burwitz2003}
\added{Burwitz, V.; Haberl, F.; Neuh\"auser, R.; Predehl, P.; Tr\"umper, J.; Zavlin, V.E. The thermal radiation of the isolated neutron star RX J1856.5-3754 observed with Chandra and XMM-Newton. {\em Astron. Astrophy.} {\bf 2003}, {\em 399}, 1109--1114.}
\bibitem[]{Li2015}
\added{Li, Z.; Qu, Z.; Chen, L.; Guo, Y.; Qu, J.; Xu, R. An Ultra-low-mass and Small-radius Compact Object in 4U 1746-37? {\em Astrophys. J.} {\bf 2015}, {\em 798}, 56.}
\bibitem[]{Yue2006}
\added{Yue, Y.L.; Cui, X.H.; Xu, R.X. Is PSR B0943+10 a Low-Mass Quark Star? {\em Astrophys. J.} {\bf 2006}, {\em 649}, L95--L98.}

\bibitem[]{Glendenning2000}
Glendenning, N.K. Compact Stars: Nuclear Physics, Particle Physics, and General Relativity, 2nd ed.; Springer-Verlag: New York, USA, 2000; pp. 337--362.
\bibitem[]{Blaschke2018}
Blaschke, D.; Chamel, N. Phases of Dense Matter in Compact Stars. In {\em The Physics and Astrophysics of Neutron Stars}; Rezzolla, L., Pizzochero, P., Jones, D.I., Rea, N., Vida\~na, I. Eds.; Springer Nature Switzerland AG: Cham, Switzerland, 2018; pp. 337--400.
\bibitem[]{Annala2020}
Annala, E.; Gorda, T.; Kurkela, A.; N\"attil\"a, J.; Vuorinen, A. Evidence for quark-matter cores in massive neutron stars. {\em Nat. Phys.} {\bf 2020}, {\em 16}, 907--910.
\bibitem[]{Berezhiani2003}
Berezhiani, Z.; Bombaci, I.; Drago, A.; Frontera, F.; Lavagno, A. Gamma-Ray Bursts from Delayed Collapse of Neutron Stars to Quark Matter Stars. {\em Astrophys. J.} {\bf 2003}, {\em 586}, 1250--1253.
\bibitem[]{Bombaci2004}
Bombaci, I.; Parenti, I.; Vida\~na, I. Quark Deconfinement and Implications for the Radius and the Limiting Mass of Compact Stars. {\em Astrophys. J.} {\bf 2004}, {\em 614}, 314--325.
\bibitem[]{Drago2004}
Drago, A.; Lavagno, A.; Pagliara, G. The Supernova-GRB connection. {\em Eur. Phys. J. A} {\bf 2004}, {\em 19}, 197--201.
\bibitem[]{Bombaci2021}
Bombaci, I.; Drago, A.; Logoteta, D.; Pagliara, G.; Vida\~na, I. Was GW190814 a Black Hole-Strange Quark Star System? {\em Phys. Rev. Lett.} {\bf 2021}, {\em 126}, 162702.
\bibitem[]{Bombaci2022}
Bombaci, I. The Equation of State of Neutron Star Matter. In {\em Millisecond Pulsars}; Bhattacharyya, S., Papitto, A., Bhattacharya, D. Eds.; Springer Nature Switzerland AG: Cham, Switzerland, 2022; pp. 281--317.
\bibitem[]{Xu2003}
Xu, R.X. Solid Quark Stars? {\em Astrophys. J. Lett.} {\bf 2003}, {\em 596}, L59. 
\bibitem[]{Miao2022a}
Miao, Z.-Q.; Xia, C.-J.; Lai, X.-Y.; Maruyama T.; Xu R.-X.; Zhou, E.-P. A bag model of matter condensed by the strong interaction. {\em Int. J. Mod. Phys. E} {\bf 2022}, {\em 31}, 2250037.
\bibitem[]{Lai2023}
Lai, X.; Xia, C.; Xu, R. Bulk strong matter: the trinity. {\em Adv. Phys. X} {\bf 2023}, {\em 8}, 2137433.
\bibitem[]{Holdom2018}
Holdom, B.; Ren, J.; Zhang, C. Quark Matter May Not Be Strange. {\em Phys. Rev. Lett.} {\bf 2018}, {\em 120}, 222001.
\bibitem[]{Zhang2019}
Zhao, T.; Zheng, W.; Wang, F.; Li, C.-M.; Yan, Y.; Huang, Y.-F.; Zong, H.-S. Do current astronomical observations exclude the existence of nonstrange quark stars? {\em Phys. Rev. D} {\bf 2019}, {\em 100}, 043018.
\bibitem[]{Zhang2020}
Zhang, C. Probing up-down quark matter via gravitational waves. {\em Phys. Rev. D} {\bf 2020}, {\em 101}, 043003.
\bibitem[]{Cao2022}
Cao, Z.; Chen, L.-W.; Chu, P.-C.; Zhou, Y. GW190814: Circumstantial evidence for up-down quark star. {\em Phys. Rev. D} {\bf 2022}, {\em 106}, 083007.


\bibitem[]{Weissenborn2011}
Weissenborn, S.; Sagert, I.; Pagliara, G.; Hempel, M.; Schaffner-Bielich, J. Quark matter in massive compact stars. {\em Astrophys. J. Lett.} {\bf 2011}, {\em 740}, L14.
\bibitem[]{Wei2012}
Wei, W.; Zheng, X.-P. Quark stars with the density-dependent quark mass model. {\em Astropart. Phys.}  {\bf 2012}, {\em 37}, 1--4.
\bibitem[]{Pi2015}
Pi, C.-M.; Yang, S.-H.; Zheng, X.-P. R-mode instability of strange stars and observations of neutron stars in LMXBs. {\em Res. Astron. Astrophy.} {\bf 2015}, {\em 15}, 871--878.
\bibitem[]{Zhou2018}
Zhou, E.-P.; Zhou, X.; Li, A. Constraints on interquark interaction parameters with GW170817 in a binary strange star scenario. {\em Phys. Rev. D} {\bf 2018}, {\em 97}, 083015.
\bibitem[]{Yang2020}
Yang, S.-H.; Pi, C.-M.; Zheng, X.-P.; Weber, F. Non-Newtonian Gravity in Strange Quark Stars and Constraints from the Observations of PSR J0740+6620 and GW170817. {\em Astrophys. J.} {\bf 2020}, {\em 902}, 32.
\bibitem[]{Yang2021a}
Yang, S.-H.; Pi, C.-M.; Zheng, X.-P. Strange stars with a mirror-dark-matter core confronting with the observations of compact stars. {\em Phys. Rev. D} {\bf 2021}, {\em 104}, 083016.
\bibitem[]{Yang2021b}
Yang, S.-H.; Pi, C.-M.; Zheng, X.-P.; Weber, F. Constraints from compact star observations on non-Newtonian gravity in strange stars based on a density dependent quark mass model. {\em Phys. Rev. D} {\bf 2021}, {\em 103}, 043012.
\bibitem[]{Cai2021}
Cai, W.-H.; Wang, Q.-W. Strange quark star and the parameter space of the quasi-particle model. {\em Commun. Theor. Phys.} {\bf 2021}, {\em 73}, 105202.
\bibitem[]{Backes2021}
Backes, B.C.; Hafemann, E.; Marzola, I.; Menezes, D.P. Density-dependent quark mass model revisited: thermodynamic consistency, stability windows and stellar properties. {\em J. Phys. G: Nucl. Part. Phys.} {\bf 2021}, {\em 48}, 055104.
\bibitem[]{Pi2022a}
Pi, C.-M.; Yang, S.-H. Non-Newtonian gravity in strange stars and constraints from the observations of compact stars. {\em New Astron.} {\bf 2022}, {\em 90}, 101670.
\bibitem[]{Pi2022b}
Pi, C.-M.; Yang, S.-H. Strange stars confronting with the observations: Non-Newtonian gravity effects, or the existence of a dark-matter core. {\em Astron. Nachr.,} {\bf 2023}, e20220083.

\bibitem[]{Demorest2010}
Demorest, P.B., Pennucci, T., Ransom, S.M., Roberts, M.S.E., Hessels, J.W.T. A two-solar-mass neutron star measured using Shapiro delay. Nature, 467, 1081--1083.
\bibitem[]{Antoniadis2013}
Antoniadis, J.; Freire, P.C.C.; Wex, N.; Tauris, T.M. ; Lynch, R.S.; van Kerkwijk, M.H.; Kramer, M.; Bassa, C.; Dhillon, V.S.; Driebe, T.; et al. A Massive Pulsar in a Compact Relativistic Binary. {\em Science} {\bf 2013}, {\em 340}, 1233232.

\bibitem[]{Abbott2017}
Abbott, B.P.; et al. (LIGO Scientific and Virgo Collaborations). GW170817: Observation of Gravitational Waves from a Binary Neutron Star Inspiral. {\em Phys. Rev. Lett.} {\bf 2017}, {\em 119}, 161101.
\bibitem[]{Abbott2018}
Abbott, B.P.; et al. (LIGO Scientific and Virgo Collaborations). GW170817: Measurements of Neutron Star Radii and Equation of State. {\em Phys. Rev. Lett.} {\bf 2018}, {\em 121}, 161101.

\bibitem[]{Cromartie2020}
Cromartie, H.T.; Fonseca, E.; Ransom, S.M.; Demorest, P.B.; Arzoumanian, Z.; Blumer, H.; Brook, P.R.; DeCesar, M.E.; Dolch, T.; Ellis, J.A.; et al. Relativistic Shapiro delay measurements of an extremely massive millisecond pulsar. {\em Nat. Astron} {\bf 2020}, {\em 4}, 72--76.
\bibitem[]{Fonseca2021}
Fonseca, E.; Cromartie, H.; Pennucci, T.T.; Ray, P.S.; Kirichenko, A.Y.; Ransom, S.M.; Demorest, P.B.; Stairs, I.H.; Arzoumanian, Z.; Guillemot, L.; et al. Refined Mass and Geometric Measurements of the High-Mass PSR J0740+6620. {\em Astrophys. J. Lett.} {\bf 2021}, {\em 915}, L12.

\bibitem[]{Doneva2018}
Doneva, D.D.; Pappas, G. Universal Relations and Alternative Gravity Theories. In {\em The Physics and Astrophysics of Neutron Stars}; Rezzolla, L., Pizzochero, P., Jones, D.I., Rea, N., Vida\~na, I. Eds.; Springer Nature Switzerland AG: Cham, Switzerland, 2018; pp. 737--806.

\bibitem[]{Krivoruchenko2009}
Krivoruchenko, M.I.; \v{S}imkovic, F.; Faessler, A. Constraints for weakly interacting light bosons from existence of massive neutron stars. {\em Phys. Rev. D} {\bf 2009}, {\em 79}, 125023.
\bibitem[]{Wen2009}
Wen, D.-H.; Li, B.-A.; Chen, L.W. Supersoft symmetry energy encountering non-Newtonian gravity in neutron stars. {\em Phys. Rev. Lett.} {\bf 2009}, {\em 103}, 211102.
\bibitem[]{Cooney2010}
Cooney, A.; Dedeo, S.; Psaltis, D. Neutron stars in f(R) gravity with perturbative constraints. {\em Phys. Rev. D} {\bf 2010}, {\em 82}, 064033.
\bibitem[]{Arapoglu2011}
Arapo\v glu, S.; Deliduman, C.; Ekşi, K.Y. Constraints on perturbative f(R) gravity via neutron stars. {J. Cosmol. Astropart. Phys.} {\bf 2011}, {\em 07}, 020. 
\bibitem[]{Pani2011a}
Pani, P.; Cardoso, V.; Delsate, T. Compact Stars in Eddington Inspired Gravity. {\em Phys. Rev. Lett} {\bf 2011}, {\em 107}, 031101.  
\bibitem[]{Pani2011b}
Pani, P; Berti, E.; Cardoso, V.; Read, J. Compact stars in alternative theories of gravity: Einstein-Dilaton-Gauss-Bonnet gravity. {\em Phys. Rev. D} {\bf 2011}, {\em 84}, 104035. 
\bibitem[]{Rahaman2012}
Rahaman, F.; Sharma, R.; Ray, S.; Maulick, R.; Karar, I. Strange stars in Krori-Barua space-time. {\em Eur. Phys. J. C} {\bf 2012}, {\em 72}, 2071.
\bibitem[]{Harko2013} 
Harko, T.; Lobo, F.S.N.; Mak, M.K.; Sushkov, S.V. Structure of neutron, quark, and exotic stars in Eddington-inspired Born-Infeld gravity. {\em Phys. Rev. D} {\bf 2013}, {\em 88}, 044032. 
\bibitem[]{Staykov2014}
Staykov, K.V.; Doneva, D.D.; Yazadjiev, S.S.; Kokkotas, K.D. Slowly rotating neutron and strange stars in R$^2$ gravity. {J. Cosmol. Astropart. Phys.} {\bf 2014}, {\em 10}, 006.    
\bibitem[]{Sham2014}
Sham, Y.-H.; Lin, L.-M.; Leung, P.T. Testing Universal Relations of Neutron Stars with a Nonlinear Matter-Gravity Coupling Theory. {\em Astrophys. J.} {\bf 2014}, {\em 781}, 66.
\bibitem[]{Moraes2016}
Moraes, P.H.R.S.; Arba\~nil, J.D.V.; Malheiro, M. Stellar equilibrium configurations of compact stars in f(R, T) theory of gravity. {J. Cosmol. Astropart. Phys.} {\bf 2016}, {\em 06}, 005. 
\bibitem[]{Yagi2017}
Yagi, K.; Yunes, N. Approximate universal relations for neutron stars and quark stars. {Phys. Rep.} {\bf 2017}, {\em 681}, 1--72. 
\bibitem[]{Yazadjiev2018}
Yazadjiev, S.S.; Doneva, D.D.; Kokkotas, K.D. Tidal Love numbers of neutron stars in f(R) gravity. {\em Eur. Phys. J. C} {\bf 2018}, {\em 78}, 818.
\bibitem[]{Lopes2018}
Lopes, I.; Panotopoulos, G. Dark matter admixed strange quark stars in the Starobinsky model. {\em Phys. Rev. D} {\bf 2018}, {\em 97}, 024030. 
\bibitem[]{Debabrata2019}
Deb, D.; Ketov, S.V.; Khlopov, M.; Ray, S. Study on charged strange stars in f(R, T) gravity. {J. Cosmol. Astropart. Phys.} {\bf 2019}, {\em 10}, 070.   
\bibitem[]{Salako2020}
Salako, I.G.; Khlopov, M.; Ray, S.; Arouko, M.Z.; Saha, P.; Debnath, U. Study on Anisotropic Strange Stars in f(T, T) Gravity. {Universe} {\bf 2020}, {\em 6}, 167.   
\bibitem[]{Majid2020}
Majid, A.; Sharif, M. Quark Stars in Massive Brans-Dicke Gravity with Tolman-Kuchowicz Spacetime. {Universe} {\bf 2020}, {\em 6}, 124.   
\bibitem[]{Astashenok2020}
Astashenok, A.V.; Capozziello, S.; Odintsov, S.D.; Oikonomou, V.K. Extended gravity description for the GW190814 supermassive neutron star. {Phys. Lett. B} {\bf 2020}, {\em 811}, 135910.  
\bibitem[]{Danchev2021}
Danchev, V.I.; Doneva, D.D. Constraining the equation of state in modified gravity via universal relations. {\em Phys. Rev. D} {\bf 2021}, {\em 103}, 024049.    
\bibitem[]{Saffer2021}
Saffer, A.; Yagi, K. Tidal deformabilities of neutron stars in scalar-Gauss-Bonnet gravity and their applications to multimessenger tests of gravity. {\em Phys. Rev. D} {\bf 2021}, {\em 104}, 124052.  
\bibitem[]{Banerjee2021}
Banerjee, A.; Tangphati, T.; Channuie, P. Strange Quark Stars in 4D Einstein-Gauss-Bonnet Gravity. {\em Astrophys. J.} {\bf 2021}, {\em 909}, 14.

\bibitem[]{Astashenok2021}
Astashenok, A.V.; Capozziello, S.; Odintsov, S.D.; Oikonomou, V.K. Maximum baryon masses for static neutron stars in f(R) gravity. {\em Europhys. Lett.} {\bf 2021}, {\em 136}, 59001.
\bibitem[]{Prasetyo2021}
Prasetyo, I.; Maulana, H.; Ramadhan, H.S.; Sulaksono, A. 2.6 $M_{\odot}$ compact object and neutron stars in Eddington-inspired Born-Infeld theory of gravity. {\em Phys. Rev. D} {\bf 2021}, {\em 104}, 084029. 
\bibitem[]{Panotopoulos2021}
Panotopoulos, G.; Tangphati, T.; Banerjee, A.; Jasim, M.K. Anisotropic quark stars in R$^2$ gravity. {Phys. Lett. B} {\bf 2021}, {\em 817}, 136330.

\bibitem[]{Xu2022}
Xu, R.; Gao, Y.; Shao, L. Neutron stars in massive scalar-Gauss-Bonnet gravity: Spherical structure and time-independent perturbations. {\em Phys. Rev. D} {\bf 2022}, {\em 105}, 024003.

\bibitem[]{Tangphati2022}
Tangphati, T.; Hansraj, S.; Banerjee, A.; Pradhan, A. Quark stars in f(R, T) gravity with an interacting quark equation of state.  {\em Phys. Dark Universe} {\bf 2022}, {\em 35}, 100990.
\bibitem[]{Lin2022}
Lin, R.-H.; Chen, X.-N.; Zhai, X.-H. Realistic neutron star models in f(T) gravity. {\em Eur. Phys. J. C} {\bf 2022}, {\em 82}, 308.
\bibitem[]{Jimenez2022a}
Jim\'enez, J.C.; Pretel, J.M.Z.; Fraga, E.S.; Jor\'as, S.E.; Reis, R.R.R. R$^2$-gravity quark stars from perturbative QCD. {J. Cosmol. Astropart. Phys.} {\bf 2022}, {\em 07}, 017.   
\bibitem[]{Pretel2022}
Pretel, J.M.Z.; Arba\~nil, J.D.V.; Duarte, S.B.; Jor\'s, S.E.; Reis, R.R.R. Charged quark stars in metric f(R) gravity. {J. Cosmol. Astropart. Phys.} {\bf 2022}, {\em 09}, 058.
\bibitem[]{Shao2022}
Shao, L.; Yagi, K. Neutron stars as extreme laboratories for gravity tests. {Sci. Bull.} {\bf 2022}, {\em 67}, 1946--1949.
\bibitem[]{Hanafy2022}
El Hanafy, W. Impact of Rastall Gravity on Mass, Radius, and Sound Speed of the Pulsar PSR J0740+6620. {\em Astrophys. J.} {\bf 2022}, {\em 940}, 51.
\bibitem[]{Yang2022c}
Yang, R.-X.; Xie, F.; Liu, D.-J. Tidal Deformability of Neutron Stars in Unimodular Gravity. {\em Universe} {\bf 2022}, {\em 8}, 576.  
\bibitem[]{Carvalho2022}
Carvalho, G.A.; Lobato, R.V.; Deb, D.; Moraes, P.H.R.S.; Malheiro, M. Quark stars with 2.6 $M_{\odot}$ in a non-minimal geometry-matter coupling theory of gravity. {\em Eur. Phys. J. C} {\bf 2022}, {\em 82}, 1096.  
\bibitem[]{Maurya2023}
Maurya, S.K.; Singh, K.N.; Govender, M.; Ray, S. Observational constraints on maximum mass limit and physical properties of anisotropic strange star models by gravitational decoupling in Einstein-Gauss-Bonnet gravity. {\em Mon. Not. R. Astron. Soc.} {\bf 2023}, {\em 519}, 4303--4324.

\bibitem[]{Shao2019}
Shao, L. Degeneracy in studying the supranuclear equation of state and modified gravity with neutron stars. {\em AIP Conf. Proc.} {\bf 2019}, {\em 2127}, 020016.
\bibitem[]{Olmo2020}
Olmo, G.J.; Rubiera-Garcia, D.; Wojnar, A. Stellar structure models in modified theories of gravity: Lessons and challenges. {\em Phys. Rep.} {\bf 2020}, {\em 876}, 1--75.

\bibitem {Spergel2000}
Spergel, D.N.; Steinhardt, P.J. Observational Evidence for Self-Interacting Cold Dark Matter. {\em Phys. Rev. Lett.} {\bf 2000}, {\em 84}, 3760.
\bibitem {Tulin2018}
Tulin, S.; Yu, H.-B. Dark matter self-interactions and small scale structure. {\em Phys. Rep.} {\bf 2018}, {\em 730}, 1--57.
\bibitem {Bertone2018}
Bertone, G.; Tait, T.M.P. A new era in the search for dark matter. {\em Nature} {\bf 2018}, {\em 562}, 51--56.
\bibitem[]{Sandin2009}
Sandin, F.; Ciarcelluti, P. Effects of mirror dark matter on neutron stars. {\em Astropart. Phys.} {\bf 2009}, {\em 32}, 278--284.
\bibitem[]{Ciarcelluti2011}
Ciarcelluti, P.; Sandin, F. Have neutron stars a dark matter core? {\em Phys. Lett. B} {\bf 2011}, {\em 695}, 19--21.
\bibitem {Leung2011}
Leung, S.-C.; Chu, M.-C.; Lin, L.-M. Dark-matter admixed neutron stars. {\em Phys. Rev. D} {\bf 2011}, {\em 84}, 107301.
\bibitem {Li2012a}
Li, A.; Huang, F.; Xu, R.-X. Too massive neutron stars: The role of dark matter? {Astropart. Phys.} {\bf 2012}, {\em 37}, 70--74. 
\bibitem {Li2012}
Li, X.Y.; Wang, F.Y.; Cheng, K.S. Gravitational effects of condensate dark matter on compact stellar objects. {J. Cosmol. Astropart. Phys.} {\bf 2012}, {\em 10}, 031. 
\bibitem {Xiang2014}
Xiang, Q.-F.; Jiang, W.-Z.; Zhang, D.-R.; Yang, R.-Y. Effects of fermionic dark matter on properties of neutron stars. {\em Phys. Rev. C} {\bf 2014}, {\em 89}, 025803. 
\bibitem {Mukhopadhyay2017} 
Mukhopadhyay, S.; Atta, D.; Imam, K.; Basu, D.N.; Samanta, C. Compact bifluid hybrid stars: hadronic matter mixed with self-interacting fermionic asymmetric dark matter. {\em Eur. Phys. J. C} {\bf 2017}, {\em 77}, 440. 
\bibitem {Ellis2018} 
Ellis, J.; H\"utsi, G.; Kannike, K.; Marzola, L.; Raidal, M.; Vaskonen, V. Dark matter effects on neutron star properties. {\em Phys. Rev. D} {\bf 2018}, {\em 97}, 123007. 
\bibitem {Deliyergiyev2019} 
Deliyergiyev, M.; Del Popolo, A.; Tolos, L.; Le Delliou, M.; Lee, X.; Burgio, F. Dark compact objects: An extensive overview. {\em Phys. Rev. D} {\bf 2019}, {\em 99}, 063015.
\bibitem {Bezares2019} 
Bezares, M.; Vigano, D.; Palenzuela, C. Gravitational wave signatures of dark matter cores in binary neutron star mergers by using numerical simulations. {\em Phys. Rev. D} {\bf 2019}, {\em 100}, 044049.
\bibitem {Ivanytskyi2020} 
Ivanytskyi, O.; Sagun, V.; Lopes, I. Neutron stars: New constraints on asymmetric dark matter. {\em Phys. Rev. D} {\bf 2020}, {\em 102}, 063028.
\bibitem[]{Kain2021}
Kain, B. Dark matter admixed neutron stars. {\em Phys. Rev. D} {\bf 2021}, {\em 103}, 043009.
\bibitem[]{Berezhiani2021}
Berezhiani, Z.; Biondi, R.; Mannarelli, M.; Tonelli, F. Neutron-mirror neutron mixing and neutron stars. {\em Eur. Phys. J. C} {\bf 2021}, {\em 81}, 1036.
\bibitem[]{Ciancarella2021}
Ciancarella, R.; Pannarale, F.; Addazi, A.; Marciano, A. Constraining mirror dark matter inside neutron stars. {\em Phys. Dark Universe} {\bf 2021}, {\em 32}, 100796.
\bibitem[]{Emma2022}
Emma, M.; Schianchi, F.; Pannarale, F.; Sagun, V.; Dietrich, T. Numerical Simulations of Dark Matter Admixed Neutron Star Binaries. {\em Particles} {\bf 2022}, {\em 5}, 273--286. 
\bibitem {Karkevandi2022} 
Rafiei Karkevandi, D.; Shakeri, S.; Sagun, V.; Ivanytskyi, O. Bosonic dark matter in neutron stars and its effect on gravitational wave signal. {\em Phys. Rev. D} {\bf 2022}, {\em 105}, 023001.
\bibitem {Giovanni2021} 
Di Giovanni, F.; Sanchis-Gual, N.; Cerd\'a-Dur\'an, P.; Font, J.A. Can fermion-boson stars reconcile multimessenger observations of compact stars? {\em Phys. Rev. D} {\bf 2022}, {\em 105}, 063005.
\bibitem {Leung2022} 
Leung, K.-L.; Chu, M.-C.; Lin, L.-M. Tidal deformability of dark matter admixed neutron stars. {\em Phys. Rev. D} {\bf 2022}, {\em 105}, 123010.
\bibitem[]{Das2022a}
Das, A.; Malik, T.; Nayak, A.C. Dark matter admixed neutron star properties in light of gravitational wave observations: A two fluid approach. {\em Phys. Rev. D} {\bf 2022}, {\em 105}, 123034.
\bibitem {Gleason2022} 
Gleason,T.; Brown, B.; Kain, B. Dynamical evolution of dark matter admixed neutron stars. {\em Phys. Rev. D} {\bf 2022}, {\em 105}, 023010.
\bibitem[]{Dengler2022}
Dengler, Y.; Schaffner-Bielich, J.; Tolos, L. Second Love number of dark compact planets and neutron stars with dark matter. {\em Phys. Rev. D} {\bf 2022}, {\em 105}, 043013.
\bibitem[]{Miao2022b}
Miao, Z.; Zhu, Y.; Li, A.; Huang, F. Dark Matter Admixed Neutron Star Properties in the Light of X-Ray Pulse Profile Observations. {\em Astrophys. J.} {\bf 2022}, {\em 936}, 69.
\bibitem[]{Collier2022} 
Collier, M.; Croon, D.; Leane, R.K. Tidal Love numbers of novel and admixed celestial objects. {\em Phys. Rev. D} {\bf 2022}, {\em 106}, 123027.
\bibitem[]{Rutherford2022}
Rutherford, N.; Raaijmakers, G.; Prescod-Weinstein, C.; Watts, A. Constraining bosonic asymmetric dark matter with neutron star mass-radius measurements. arXiv 2022, arXiv:2208.03282.
\bibitem[]{Giangrandi2022}
Giangrandi, E.; Sagun, V.; Ivanytskyi, O.; Provid\^encia, C.; Dietrich, T. The effects of self-interacting bosonic dark matter on neutron star properties. arXiv 2022, arXiv:2209.10905.
\bibitem[]{Shakeri2022}
Shakeri, S.; Rafiei Karkevandi, D. Bosonic Dark Matter in Light of the NICER Precise Mass-Radius Measurements. arXiv 2022, arXiv:2210.17308.
\bibitem[]{Hippert2022}
Hippert, M.; Dillingham, E.; Tan, H.; Curtin, D.; Noronha-Hostler, J.; Yunes, N. Dark Matter or Regular Matter in Neutron Stars? How to tell the difference from the coalescence of compact objects. arXiv 2022, arXiv:2211.08590.
\bibitem[]{Fynn2023}
\added{Fynn Diedrichs, R.; Becker, N.; Jockel, C.; Christian, J.-E.; Sagunski, L.; Schaffner-Bielich, J. Tidal Deformability of Fermion-Boson Stars: Neutron Stars Admixed with Ultra-Light Dark Matter. arXiv 2023, arXiv: 2303.04089.}
\bibitem {Mukhopadhyay2016}   
Mukhopadhyay, P.; Schaffner-Bielich, J. Quark stars admixed with dark matter. {\em Phys. Rev. D} {\bf 2016}, {\em 93}, 083009.
\bibitem {Panotopoulos2017} 
Panotopoulos, G.; Lopes, I. Gravitational effects of condensed dark matter on strange stars. {\em Phys. Rev. D} {\bf 2017}, {\em 96}, 023002.
\bibitem {Panotopoulos2018} 
Panotopoulos, G.; Lopes, I. Radial oscillations of strange quark stars admixed with fermionic dark matter. {\em Phys. Rev. D} {\bf 2018}, {\em 98}, 083001. 
\bibitem {Jimenez2022b} 
 Jim\'{e}nez, J.C.; Fraga, E.S. Radial Oscillations of Quark Stars Admixed with Dark Matter. {\em Universe} {\bf 2022}, {\em 8}, 34.
\bibitem {Ferreira2022} 
Ferreira, O.; Fraga, E.S. Strange magnetars admixed with fermionic dark matter. {J. Cosmol. Astropart. Phys.} {\bf 2023}, {\em 04}, 012. 
\bibitem[]{Lopes2023}
Lopes,L.L.; Das, H. C. Strange Stars within Bosonic and Fermionic Admixed Dark Matter. arXiv 2023. arXiv:2301.00567.

\bibitem[]{Zyla2020}
Zyla, P.A.; et al. (Particle Data Group). Review of Particle Physics. {\em Prog. Theor. Exp. Phys.} {\bf 2020}, {\em 2020}, 083C01.
\bibitem[]{Tolman1939}
Tolman, R.C. Static Solutions of Einstein's Field Equations for Spheres of Fluid. {\em Phys. Rev.} {\bf 1939}, {\em 55}, 364--373.
\bibitem[]{Oppenheimer1939}
Oppenheimer, J.R.; Volkoff, G.M. On Massive Neutron Cores. {\em Phys. Rev.} {\bf 1939}, {\em 55}, 374--381.
\bibitem[]{Flanagan2008}
Flanagan, \'E.\'E.; Hinderer, T. Constraining neutron-star tidal Love numbers with gravitational-wave detectors. {\em Phys. Rev. D} {\bf 2008}, {\em 77}, 021502.
\bibitem[]{Hinderer2008}
Hinderer, T. Tidal Love Numbers of Neutron Stars. {\em Astrophys. J.} {\bf 2008}, {\em 677}, 1216--1220.
\bibitem[]{Damour2009}
Damour, T.; Nagar, A. Relativistic tidal properties of neutron stars. {\em Phys. Rev. D} {\bf 2009}, {\em 80}, 084035.
\bibitem[]{Hinderer2010}
Hinderer, T.; Lackey, B.D.; Lang, R.N.; Read, J.S. Tidal deformability of neutron stars with realistic equations of state and their gravitational wave signatures in binary inspiral. {\em Phys. Rev. D} {\bf 2010}, {\em 81}, 123016.
\bibitem[]{Postnikov2010}
Postnikov, S.; Prakash, M.; Lattimer, J.M. Tidal Love numbers of neutron and self-bound quark stars. {\em Phys. Rev. D} {\bf 2010}, {\em 82}, 024016.

\bibitem[]{Capano2020}
Capano, C.D.; Tews, I.; Brown, S.M.; Margalit, B.; De, S.; Kumar, S.; Brown, D.A.; Krishnan, B.; Reddy, S. Stringent constraints on neutron-star radii from multimessenger observations and nuclear theory. {\em Nat. Astron.} {\bf 2020}, {\em 4}, 625--632.
\bibitem[]{Riley2019}
Riley, T.E.; Watts, A.L.; Bogdanov, S.; Ray, P.S.; Ludlam, R.M.; Guillot, S.; Arzoumanian, Z.; Baker, C.L.; Bilous, A.V.; Chakrabarty, D.; et al. A NICER View of PSR J0030+0451: Millisecond Pulsar Parameter Estimation. {\em Astrophys. J. Lett.} {\bf 2019}, {\em 887}, L21. 
\bibitem[]{Miller2019}
Miller, M.C.; Lamb, F.K.; Dittmann, A.J.; Bogdanov, S.; Arzoumanian, Z.; Gendreau, K.C.; Guillot, S.; Harding, A.K.; Ho, W.C.G.; Lattimer. J.M.; et al. PSR J0030+0451 Mass and Radius from NICER Data and Implications for the Properties of Neutron Star Matter. {\em Astrophys. J. Lett.} {\bf 2019}, {\em 887}, L24. 
\bibitem[]{Riley2021}
Riley, T.E.; Watts, A.L.; Ray, P.S.; Bogdanov, S.; Guillot, S.; Morsink, S.M.; Bilous, A.V.; Arzoumanian, Z.; Choudhury, D.; Deneva, J.S.; et al. A NICER View of the Massive Pulsar PSR J0740+6620 Informed by Radio Timing and XMM-Newton Spectroscopy. {\em Astrophys. J. Lett.} {\bf 2021}, {\em 918}, L27. 
\bibitem[]{Miller2021}
Miller, M.C.; Lamb, F.K.; Dittmann, A.J.; Bogdanov, S.; Arzoumanian, Z.; Gendreau, K.C.; Guillot, S.; Ho,W.C.G.; Lattimer, J.M.; Loewenstein, M.; et al. The Radius of PSR J0740+6620 from NICER and XMM-Newton Data. {\em Astrophys. J. Lett.} {\bf 2021}, {\em 918}, L28. 

\bibitem[]{Schaab1997}
Schaab, C.; Hermann, B.; Weber, F.; Weigel, M.K. Are strange stars distinguishable from neutron stars by their cooling behaviour? {\em J. Phys. G: Nucl. Part. Phys.} {\bf 1997}, {\em 23}, 2029--2037.

\bibitem[]{Alford2001}
\added{Alford, M.; Rajagopal, K.; Reddy, S.; Wilczek, F. Minimal color-flavor-locked-nuclear interface. {\em Phys. Rev. D} {\bf 2001}, {\em 64}, 074017.}
\bibitem[]{Oertel2008}
\added{Oertel, M.; Urban, M. Surface effects in color superconducting strange-quark matter. {\em Phys. Rev. D} {\bf 2008}, {\em 77}, 074015.} 
\bibitem[]{Lugones2013}
\added{Lugones, G.; Grunfeld, A.; Ajmi, M.A. Surface tension and curvature energy of quark matter in the Nambu-Jona-Lasinio model. {\em Phys. Rev. C} {\bf 2013}, {\em 88}, 045803.}
\bibitem[]{Xia2013}
\added{Xia, C.J.; Peng, G.X.; Sun, T.T.; Guo, W.L.; Lu, D.H.; Jaikumar, P. Interface effects of strange quark matter with density dependent quark masses. {\em Phys. Rev. D} {\bf 2018}, {\em 98}, 034031.} 
\bibitem[]{Fraga2019}
\added{Fraga, E.S.; Hippert, M.; Schmitt, A. Surface tension of dense matter at the chiral phase transition. {\em Phys. Rev. D} {\bf 2019}, {\em 99}, 014046.} 
\bibitem[]{Lugones2019}
\added{Lugones, G.; Grunfeld, A.G. Surface tension of hot and dense quark matter under strong magnetic fields. {\em Phys. Rev. C} {\bf 2019}, {\em 99}, 035804.}
\bibitem[]{Wang2021}
\added{Wang, L.; Hu, J.; Xia, C.-J.; Xu, J.-F.; Peng, G.-X.; Xu, R.-X. Stable Up-Down Quark Matter Nuggets, Quark Star Crusts, and a New Family of White Dwarfs. {\em Galaxies} {\bf 2021}, {\em 9}, 70.}


\bibitem[]{Romani2022}
Romani, R.W.; Kandel, D.; Filippenko, A.V.; Brink, T.G.; Zheng, W. PSR J0952-0607: The Fastest and Heaviest Known Galactic Neutron Star. {\em Astrophys. J. Lett.} {\bf 2022}, {\em 934}, L17. 
\bibitem[]{Konstantinou2022}
Konstantinou, A.; Morsink, S.M. Universal Relations for the Increase in the Mass and Radius of a Rotating Neutron Star. {\em Astrophys. J.} {\bf 2022}, {\em 934}, 139.

\bibitem[]{Adelberger2009}
Adelberger, E.G.; Gundlach, J.H.; Heckel, B.R.; Hoedl, S.; Schlamminger, S. Torsion balance experiments: a low-energy frontier of particle physics. {\em  Prog. Part. Nucl. Phys.} {\bf 2009}, {\em 62}, 102--134.
\bibitem[]{Adelberger2003}
Adelberger, E.G.; Heckel, B.R.; Nelson, A.E. Tests of the gravitational inverse-square law. {\em Annu. Rev. Nucl. Part. Sci.} {\bf 2003}, {\em 53}, 77--121.
\bibitem[]{Fischbach1999}
Fischbach, E.; Talmadge, C.L. \textit{The Search for Non-Newtonian Gravity}; Springer-Verlag: New York, NY, USA, 1999.
\bibitem[]{Fayet1980}
Fayet, P. Effects of the spin-1 partner of the goldstino (gravitino) on neutral current phenomenology. {\em Phys. Lett. B} {\bf 1980}, {\em 95}, 285--289.
\bibitem[]{Fayet1981}
Fayet, P. A la recherche d'un nouveau boson de spin un. {\em Nucl. Phys. B} {\bf 1980}, {\em 187}, 184--204.
\bibitem[]{Murata2015}
Murata, J.; Tanaka, S. A review of short-range gravity experiments in the LHC era. {\em Class. Quant. Grav.} {\bf 2015}, {\em 32}, 033001.
\bibitem[]{Wen2011}
Wen, D.-H.; Li, B.-A.; Chen, L.-W. Can the maximum mass of neutron stars rule out any equation of state of dense stellar matter before gravity is well understood? arXiv 2011, arXiv:1101.1504.
\bibitem[]{Sulaksono2011}
Sulaksono, A.; Marliana; Kasmudin. Effects of In-Medium Modification of Weakly Interacting Light Boson Mass in Neutron Stars. {\em Mod. Phys. Lett. A} {\bf 2011}, {\em 26}, 367--375. 
\bibitem[]{Zhang2011}
Zhang, D.-R.; Yin, P.-L; Wang, W.; Wang, Q.-C.; Jiang, W.-Z. Effects of a weakly interacting light U boson on the nuclear equation of state and properties of neutron stars in relativistic models. {\em Phys. Rev. C} {\bf 2011}, {\em 83}, 035801. 
\bibitem[]{Yan2013}
Wen, D.-H.; Yan, J. R-mode Instability of Neutron Star with Non-Newtonian Gravity. {\em  Commun. Theor. Phys.} {\bf 2013}, {\em 59}, 47--52.
\bibitem[]{Lin14}
Lin, W.; Li, B.-A.; Chen, L.-W.; Wen, D.-H.; Xu, J. Breaking the EOS-gravity degeneracy with masses and pulsating frequencies of neutron stars. {\em J. Phys. G: Nucl. Part. Phys.} {\bf 2014}, {\em 41}, 075203.
\bibitem[]{Lu2017}
Lu, Z.-Y.; Peng, G.-X.; Zhou, K. Effects of non-Newtonian gravity on the properties of strange stars. {\em Res. Astron. Astrophys.} {\bf 2017}, {\em 17}, 11.
\bibitem[]{Yu2018}
Yu, Z.;  Xu, Y.; Zhang, G.-Q.; Hu, T.-P. Effects of a Weakly Interacting Light U Boson on Protoneutron Stars Including the Hyperon-Hyperon Interactions. {\em Commun. Theor. Phys.} {\bf 2018}, {\em 69}, 417--424.

\bibitem[]{Fujii1971}
Fujii, Y. Dilaton and Possible Non-Newtonian Gravity. {\em Nat. Phys. Sci.} {\bf 1971}, {\em 234}, 5--7.
\bibitem[]{Kamyshkov2008}
Kamyshkov, Y.; Tithof, J.; Vysotsky, M. Bounds on new light particles from high-energy and very small momentum transfer np elastic scattering data. {\em Phys. Rev. D} {\bf 2008}, {\em 78}, 114029.
\bibitem[]{Xu2013}
Xu, J.; Li, B.-A.; Chen, L.-W.; Zheng, H. Nuclear constraints on non-Newtonian gravity at femtometer scale. {\em J. Phys. G: Nucl. Part. Phys.} {\bf 2013}, {\em 40}, 035107.
\bibitem[]{Pokotilovski2006}
Pokotilovski, Y. N. Constraints on new interactions from neutron scattering experiments. {\em Phys. At. Nucl.} {\bf 2006}, {\em 69}, 924--931.   
\bibitem[]{Klimchitskaya2020}
Klimchitskaya, G.L.; Kuusk, P.; Mostepanenko, V.M. Constraints on non-Newtonian gravity and axionlike particles from measuring the Casimir force in nanometer separation range. {\em Phys. Rev. D} {\bf 2020}, {\em 101}, 056013.
\bibitem[]{Kamiya2015}
Kamiya, Y.; Itagaki, K.; Tani, M.; Kim, G.N.; Komamiya, S. Constraints on New Gravitylike Forces in the Nanometer Range. {\em  Phys. Rev. Lett.} {\bf 2015}, {\em 114}, 161101.
\bibitem[]{Chen2016}
Chen, Y.-J.; Tham, W.K. ; Krause, D.E.; L\'opez, D.; Fischbach, E.; Decca, R.S. Stronger Limits on Hypothetical Yukawa Interactions in the 30-8000 nm Range. {\em  Phys. Rev. Lett.} {\bf 2016}, {\em 116}, 221102.
\bibitem[]{Yong2013}
Yong, G.-C.; Li, B.-A. Effects of nuclear symmetry energy on $\eta$ meson production and its rare decay to the dark U-boson in heavy-ion reactions. {\em Phys. Lett. B} {\bf 2013}, {\em 723}, 388--392.
\bibitem[]{Jean2003}
Jean, P.; Kn\"odlseder, J.; Lonjou, V.; Allain, M.; Roques, J.-P.; Skinner, G.K.; Teegarden, B.J.; Vedrenne, G.; von Ballmoos, P.; Cordier, B.; et al. Early SPI/INTEGRAL measurements of 511 keV line emission from the 4th quadrant of the Galaxy. {\em Astron. Astrophys.} {\bf 2003}, {\em 407}, L55-L58.
\bibitem[]{Boehm2004a}
Boehm, C.; Fayet, P.; Silk, J. Light and heavy dark matter particles. {\em Phys. Rev. D} {\bf 2004}, {\em 69}, 101302.
\bibitem[]{Boehm2004b}
Boehm, C.; Hooper, D.; Silk, J.; Casse, M.; Paul, J. MeV Dark Matter: Has It Been Detected? {\em Phys. Rev. Lett.} {\bf 2004}, {\em 92}, 101301.

\bibitem[]{Long2003}
Long, J.C.; Chan, H.W.; Churnside, A.B.; Gulbis, E.A.; Varney, M.C.M.; Price, J.C. Upper limits to submillimetre-range forces from extra space-time dimensions. {\em Nature} {\bf 2003}, {\em 421}, 922--925.
\bibitem[]{Fujii1988}
Fujii,Y. Cosmological Implications of the Fifth Force. In Proceedings of the 130th Symposium of the International Astronomical Union, Balatonfured, Hungary, June 15-20, 1987.
\bibitem[]{Abbott2020}
\added{Abbott, R. et al. (LIGO Scientific and Virgo Collaborations). GW190814: Gravitational Waves from the Coalescence of a 23 Solar Mass Black Hole with a 2.6 Solar Mass Compact Object. {\em Astrophys. J. Lett.} {\bf 2020}, {\em 896}, L44.}
\bibitem[]{Zhou2019}
\added{Zhou, E.; Tsokaros, A.; Ury\=u, K.; Xu, R.; Shibata, M. Differentially rotating strange star in general relativity. {\em Phys. Rev. D} {\bf 2019}, {\em 100}, 043015.}

\bibitem[]{Foot1991} 
Foot, R.; Lew, H.; Volkas, R.R. A model with fundamental improper spacetime symmetries. {\em Phys. Lett. B} {\bf 1991}, {\em 272}, 67--70.
\bibitem[]{Lee1956} 
Lee, T.D.; Yang, C.N. Question of Parity Conservation in Weak Interactions. {\em Phys. Rev.} {\bf 1956}, {\em 104}, 254--258. 
\bibitem[]{Foot2004} 
Foot, R. Mirror Matter-Type Dark Matter. {\em Int. J. Mod. Phys. D} {\bf 2004}, {\em 13}, 2161--2192.
\bibitem[]{Berezhiani2004} 
Berezhiani, Z. Mirror World and its Cosmological Consequences. {\em Int. J. Mod. Phys. A} {\bf 2004}, {\em 19}, 3775--3806.
\bibitem[]{Berezhiani2005} 
Berezhiani, Z. Through the Looking-Glass Alice's Adventures in Mirror World. In {\em From Fields to Strings: Circumnavigating Theoretical Physics}; Shifman M., et al., Eds.; World Scientific: Singapore, 2005; pp. 2147--2195.
\bibitem[]{Okun2007}
Okun, L.B. Mirror particles and mirror matter: 50 years of speculation and searching. {\em Phys. Usp.} {\bf 2007}, {\em 50}, 380--389.
\bibitem[]{Foot2014}
Foot, R. Mirror dark matter: Cosmology, galaxy structure and direct detection. {\em Int. J. Mod. Phys. A} {\bf 2014}, {\em 29}, 1430013.
\bibitem[]{Pavsic1974} 
Pavsic, M. External inversion, internal inversion, and reflection invariance. {\em Int. J. Theor. Phys.} {\bf 1974}, {\em 9}, 229--244.
\bibitem[]{Berezhiani2006}
Berezhiani, Z.;  Bento, L. Neutron-Mirror-Neutron Oscillations: How Fast Might They Be? {\em Phys. Rev. Lett.} {\bf 2006}, {\em 96}, 081801. 
\bibitem[]{Berezhiani2009}
Berezhiani, Z. More about neutron-mirror neutron oscillation. {\em Eur. Phys. J. C} {\bf 2009}, {\em 64}, 421--431.
\bibitem[]{Goldman2019}
Goldman, I.; Mohapatra, R.N.; Nussinov, S. Bounds on neutron-mirror neutron mixing from pulsar timing. {\em Phys. Rev. D} {\bf 2019}, {\em 100}, 123021.
\bibitem[]{McKeen2021}
McKeen, D.; Pospelov, M.; Raj, N. Neutron Star Internal Heating Constraints on Mirror Matter. {\em Phys. Rev. Lett.} {\bf 2021}, {\em 127}, 061805.
\bibitem[]{Goldman2022}
Goldman, I.; Mohapatra, R.N. ; Nussinov, S.; Zhang, Y. Neutron-Mirror-Neutron Oscillation and Neutron Star Cooling. {\em Phys. Rev. Lett.} {\bf 2022}, {\em 129}, 061103. 


\bibitem[]{Chakrabarty1989}
\added{Chakrabarty, S.; Raha, S.; Sinha, B. Strange quark matter and the mechanism of confinement. {\em Phys. Lett. B} {\bf 1989}, {\em 229}, 112--116.}
\bibitem[]{Benvenuto1995}
\added{Benvenuto, O.G.; Lugones, G. Strange matter equation of state in the quark mass-density-dependent model. {\em Phys. Rev. D} {\bf 1995}, {\em 51}, 1989--1993.}
\bibitem[]{Lugones1995}
\added{Lugones, G.; Benvenuto, O.G. Strange matter equation of state and the combustion of nuclear matter into strange matter in the quark mass-density-dependent model at T>0. {\em Phys. Rev. D} {\bf 1995}, {\em 52}, 1276--1280.}
\bibitem[]{Peng2000}
\added{Peng, G.X.; Chiang, H.C.; Zou, B.S.; Ning, P.Z.; Luo, S.J. Thermodynamics, strange quark matter, and strange stars. {\em Phys. Rev. C} {\bf 2000}, {\em 62}, 025801.}
\bibitem[]{Xia2014}
\added{Xia, C.J.; Peng, G.X.; Chen, S.W.; Lu, Z.Y.; Xu, J.F. Thermodynamic consistency, quark mass scaling, and properties of strange matter. {\em Phys. Rev. D} {\bf 2014}, {\em 89}, 105027.}

\bibitem[]{Peshier1994}
\added{Peshier, A.; Kämpfer, B.; Pavlenko, O.P.; Soff, G. An effective model of the quark-gluon plasma with thermal parton masses. {\em Phys. Lett. B} {\bf 1994}, {\em 337}, 235--239.}   
\bibitem[]{Gorenstein1995}
\added{Gorenstein, M.I.; Yang, S.N. Gluon plasma with a medium-dependent dispersion relation. {\em Phys. Rev. D} {\bf 1995}, {\em 52}, 5206--5212.}
\bibitem[]{Peshier2000}
\added{Peshier, A.; K\"ampfer, B.; Soff, G. Equation of state of deconfined matter at finite chemical potential in a quasiparticle description. {\em Phys. Rev. C} {\bf 2000}, {\em 61}, 045203.}
\bibitem[]{Zong2008}
\added{Zong, H.-S.; Sun, W.-M. A Model Study of the Equation of State of QCD. {\em Int. J. Mod. Phys. A} {\bf 2008}, {\em 23}, 3591--3612.}
\bibitem[]{Li2019c}
\added{Li, B.-L.; Cui, Z.-F.; Yu, Z.-H.; Yan, Y.; An, S.; Zong, H.-S. Structures of the strange quark stars within a quasiparticle model. {\em Phys. Rev. D} {\bf 2019}, {\em 99}, 043001.}

\bibitem[]{Ruster2004}
\added{R\"uster, S.B.; Rischke, D.H. Effect of color superconductivity on the mass and radius of a quark star. {\em Phys. Rev. D} {\bf 2004}, {\em 69}, 045011.}
\bibitem[]{Buballa2005}
\added{Buballa, M. NJL-model analysis of dense quark matter. {\em Phys. Rep.} {\bf 2005}, {\em 407}, 205--376.}
\bibitem[]{Menezes2006}
\added{Menezes, D.P.; Provid\^encia, C.; Melrose, D.B. Quark stars within relativistic models. {\em J. Phys. G: Nucl. Part. Phys.} {\bf 2006}, {\em 32}, 1081--1095.}
\bibitem[]{Klahn2013}
\added{Kl\"ahn, T.; Lastowiecki, R.; Blaschke, D. Implications of the measurement of pulsars with two solar masses for quark matter in compact stars and heavy-ion collisions: A Nambu-Jona-Lasinio model case study. {\em Phys. Rev. D} {\bf 2013}, {\em 88}, 085001.}
\bibitem[]{Chu2016}
\added{Chu, P.-C.; Wang, B.; Ma, H.-Y.; Dong, Y.-M.; Chang, S.-L.; Zheng, C.-H.; Liu, J.-T.; Zhang, X.-M. Quark matter in an SU(3) Nambu-Jona-Lasinio model with two types of vector interactions. {\em Phys. Rev. D} {\bf 2016}, {\em 93}, 094032.}

\bibitem[]{Freedman1977}
\added{Freedman, B.; Mclerran, L. Fermions and gauge vector mesons at finite temperature and density. III. The ground-state energy of a relativistic quark gas. {\em Phys. Rev. D} {\bf 1977}, {\em 16}, 1169--1185.}
\bibitem[]{Freedman1978}
\added{Freedman, B.; Mclerran, L. Quark star phenomenology. {\em Phys. Rev. D} {\bf 1978}, {\em 17}, 1109--1122.}
\bibitem[]{Kurkela2010}
\added{Kurkela, A.; Romatschke, P.; Vuorinen, A. Cold quark matter. {\em Phys. Rev. D} {\bf 2010}, {\em 81}, 105021.}

\bibitem[]{Chu2014}
\added{Chu, P.-C.; Chen, L.-W. Quark Matter Symmetry Energy and Quark Stars. {\em Astrophys. J.} {\bf 2014}, {\em 780}, 135.}
\bibitem[]{Liu2016}
\added{Liu, H.; Xu, J.; Chen, L.-W.; Sun, K.-J. Isospin properties in quark matter and quark stars within isospin-dependent quark mass models. {\em Phys. Rev. D} {\bf 2016}, {\em 94}, 065032.}
\bibitem[]{Chen2017}
\added{Chen L. Symmetry Energy in Nucleon and Quark Matter. {\em Nucl. Phys. Rev.} {\bf 2017}, {\em 34}, 20--28.}
\bibitem[]{Chu2019}
\added{Chu, P.-C.; Zhou, Y.; Qi, X.; Li, X.-H.; Zhang, Z.; Zhou, Y. Isospin properties in quark matter and quark stars within isospin-dependent quark mass models. {\em Phys. Rev. C} {\bf 2019}, {\em 99}, 035802.}


\bibitem[]{Alford1998}
\added{Alford, M.; Rajagopal, K.; Wilczek, F. QCD at finite baryon density: nucleon droplets and color superconductivity. {\em Phys. Lett. B} {\bf 1998}, {\em 422}, 247--256.}

\bibitem[]{Rapp1998}
\added{Rapp, R.; Sch\"afer, T.; Shuryak, E.V.; Velkovsky, M. Diquark Bose Condensates in High Density Matter and Instantons. {\em Phys. Rev. Lett.} {\bf 1998}, {\em 81}, 53--56.}

\bibitem{Rapp2000}
\added{Rapp, R.; Sch\"afer, T.; Shuryak, E.V.; Velkovsky, M. High-Density QCD and Instantons. {\em Ann. Phys.} {\bf 2000}, {\em 280}, 35--99.}

\bibitem{Rajagopal2001b}
\added{Rajagopal, K.; Wilczek, F. The Condensed Matter Physics of QCD, In {\em At the Frontier of Particle Physics: Handbook of QCD}; Shifman, M. Eds.; World Scientific: Singapore, {\bf 2001}; pp. 2061--2151.}

\bibitem{Alford2001c}
\added{Alford, M. Color-Superconducting Quark Matter. {\em Ann. Rev. Nucl. Part. Sci.} {\bf 2001}, {\em 51}, 131--160.}

\bibitem[]{Alford2008}
\added{Alford, M.G.; Schmitt, A.; Rajagopal, K.; Sch\"afer, T. Color superconductivity in dense quark matter. {\em Rev. Mod. Phys.} {\bf 2008}, {\em 80}, 1455--1515.}

\bibitem[]{Alford1999}
\added{Alford, M.; Rajagopal, K.; Wilczek, F. Color-flavor locking and chiral symmetry breaking in high density QCD. {\em Nucl. Phys. B} {\bf 1999}, {\em 537}, 443--458.}

\bibitem[]{Rajagopal2001}
\added{Rajagopal, K.; Wilczek, F. Enforced Electrical Neutrality of the Color-Flavor Locked Phase. {\em Phys. Rev. Lett.} {\bf 2001}, {\em 86}, 3492--3495.}

\bibitem[]{Lugones2002}
\added{Lugones, G.; Horvath, J.E. Color-flavor locked strange matter. {\em Phys. Rev. D} {\bf 2002}, {\em 66}, 074017.} 

\bibitem[]{Roupas2021}
\added{Roupas, Z.; Panotopoulos, G.; Lopes, I. QCD color superconductivity in compact stars: Color-flavor locked quark star candidate for the gravitational-wave signal GW190814. {\em Phys. Rev. D} {\bf 2021}, {\em 103}, 083015.}

\bibitem[]{Horvath2021}
\added{Horvath, J.E. ; Moraes, P.H.R.S. Modeling a 2.5 $M_{\odot}$ compact star with quark matter. {\em Int. J. Mod. Phys. D} {\bf 2021}, {\em 30}, 2150016.}

\bibitem[]{Miao2021c}
\added{Miao, Z.; Jiang, J.-L.; Li, A.; Chen, L.-W. Bayesian Inference of Strange Star Equation of State Using the GW170817 and GW190425 Data. {\em Astrophys. J. Lett.} {\bf 2021}, {\em 917}, L22.}
\bibitem[]{Pagliara2013}
\added{Pagliara, G.; Herzog, M.; R\"opke, F.K. Combustion of a neutron star into a strange quark star: The neutrino signal. {\em Phys. Rev. D} {\bf 2013}, {\em 87}, 103007.}

\bibitem[]{Geng2021}
\added{Geng, J.; Li, B.; Huang, Y. Repeating fast radio bursts from collapses of the crust of a strange star. {\em Innovation} {\bf 2021}, {\em 2}, 100152.}
\bibitem[]{Kuerban2020}
\added{Kuerban, A.; Geng, J.-J.; Huang, Y.-F.; Zong, H.-S.; Gong, H. Close-in Exoplanets as Candidates for Strange Quark Matter Objects. {\em Astrophys. J.} {\bf 2020}, {\em 890}, 41.}







%




\end{thebibliography}
\end{document}